\documentclass[floatfix,preprint,amsmath,amssymb,onecolumn,superscriptaddress,showpacs,longbibliography,12pt]{revtex4-2}
\usepackage{amsmath,amssymb}

\usepackage{graphicx}
\usepackage{bm}
\usepackage[T1]{fontenc}
\usepackage[utf8]{inputenc}

\DeclareUnicodeCharacter{0327}{\k{a}}
\DeclareUnicodeCharacter{0229}{\.{z}}
\DeclareUnicodeCharacter{2009}{\,}
\DeclareUnicodeCharacter{03B7}{$\eta$}
\DeclareUnicodeCharacter{03B4}{$\delta$}

\usepackage[colorlinks=true, citecolor=red, linkcolor=blue, urlcolor=black]{hyperref}
\usepackage[english]{babel}
\usepackage{braket}
\usepackage{printlen}
\usepackage{layouts}
\usepackage[nolist,nohyperlinks]{acronym}
\begin{acronym}[]
\acro{gs}[GS]{ground state}
\acro{ed}[ED]{exact diagonalization}
\acro{lmc}[LMC]{light-matter coupling}
\acro{usc}[USC]{ultra-strong coupling}
\acro{sc}[SC]{strong coupling}
\end{acronym}

\setcitestyle{super}

\begin{document}

\title{Cavity engineering of Hubbard $U$ via phonon polaritons}

\author{Brieuc Le D\'e}
\thanks{These authors contributed equally.}
\affiliation 
{Max Planck Institute for the Structure and Dynamics of Matter, Center for Free-Electron Laser Science, Luruper Chaussee 149, 22761 Hamburg, Germany}

\author{Christian J.~Eckhardt}
\thanks{These authors contributed equally.}
\affiliation 
{Institut f\"ur Theorie der Statistischen Physik, RWTH Aachen University and JARA-Fundamentals of Future Information Technology, 52056 Aachen, Germany}
\affiliation 
{Max Planck Institute for the Structure and Dynamics of Matter, Center for Free-Electron Laser Science,  Luruper Chaussee 149, 22761 Hamburg, Germany}

\author{Dante M.~Kennes}
\affiliation 
{Institut f\"ur Theorie der Statistischen Physik, RWTH Aachen University and JARA-Fundamentals of Future Information Technology, 52056 Aachen, Germany}
\affiliation 
{Max Planck Institute for the Structure and Dynamics of Matter, Center for Free-Electron Laser Science,  Luruper Chaussee 149, 22761 Hamburg, Germany}

\author{Michael A.~Sentef}
\email{michael.sentef@mpsd.mpg.de}
\affiliation 
{Max Planck Institute for the Structure and Dynamics of Matter, Center for Free-Electron Laser Science,  Luruper Chaussee 149, 22761 Hamburg, Germany}

\begin{abstract}
Pump-probe experiments have suggested the possibility to control electronic correlations by driving infrared-active phonons with resonant midinfrared laser pulses.
In this work we study two possible microscopic nonlinear electron-phonon interactions behind these observations, namely coupling of the squared lattice displacement either to the electronic density or to the double occupancy. We investigate whether photon-phonon coupling to quantized light in an optical cavity enables similar control over electronic correlations.
We first show that inside a dark cavity electronic interactions increase, ruling out the possibility that $T_c$ in superconductors can be enhanced via effectively decreased electron-electron repulsion through nonlinear electron-phonon coupling in a cavity.
We further find that upon driving the cavity, electronic interactions decrease. 
Two different regimes emerge: (i) a strong coupling regime where the phonons show a delayed response at a time proportional to the inverse coupling strength, and (ii) an ultra-strong coupling regime where the response is immediate when driving the phonon polaritons resonantly.
We further identify a distinctive feature in the electronic spectral function when electrons couple to phonon polaritons involving an infrared-active phonon mode, namely the splitting of the shake-off band into three bands.
This could potentially be observed by angle-resolved photoemission spectroscopy.
\end{abstract}

\maketitle

\section{Introduction}
The ultrafast optical control of nonthermal phases of matter in quantum materials is a blossoming research field.\cite{de_la_torre_colloquium_2021, disa_engineering_2021} Among the most intriguing experimental results are reports that suggest the possibility to induce transient superconducting-like states through laser driving. \cite{cavalleri_photo-induced_2018, nicoletti_optically_2014, cremin_photoenhanced_2019, rajasekaran_probing_2018, zhang_light-induced_2018, suzuki_photoinduced_2019, buzzi_photomolecular_2020, mitrano_possible_2016, mankowsky_nonlinear_2014, hu_optically_2014, fausti_light-induced_2011, cantaluppi_pressure_2018, liu_pump_2020}
This effect was observed in several classes of materials that share the common feature of a superconducting \ac{gs}, implying the interpretation that the laser driving effectively raises the material's critical temperature.
In order to explain the observed behaviour a number of different microscopic mechanisms were subsequently proposed. \cite{kennes_transient_2017, knap_dynamical_2016, sentef_theory_2017, kaiser_optical_2014, kaiser_optically_2014, singla_thz-frequency_2015, nava_cooling_2018, coulthard_enhancement_2017, bittner_possible_2019, dolgirev_periodic_2021, ido_correlation-induced_2017, kim_enhancing_2016, sentef_theory_2016, schutt_controlling_2018, tindall_heating-induced_2019, sun_transient_2020, tindall_dynamical_2020, dehghani_optical_2020, patel_light-induced_2016, robertson_nonequilibrium_2009, tikhonov_superconductivity_2018, li_-paired_2020, hoppner_redistribution_2015, denny_proposed_2015, komnik_bcs_2016, murakami_nonequilibrium_2017, babadi_theory_2017, michael_parametric_2020, buzzi_higgs-mediated_2021, dai_superconductinglike_2021, peronaci_enhancement_2020, lemonik_quench_2018, okamoto_theory_2016, raines_enhancement_2015, okamoto_transiently_2017, schlawin_terahertz_2017}
However, to date no final and unifying conclusion could be drawn, neither on the nature of the transient states nor on the mechanism behind them.

From a practical point of view, a drawback of the transient superconducting-like states is their relatively short life time, typically in the picosecond range, with a recent extension to the nanosecond regime in K$_3$C$_{60}$.\cite{budden_evidence_2021}
As an alternative route to control over material properties, light-matter coupling in cavities has been suggested.\cite{ruggenthaler_quantum-electrodynamical_2018, frisk_kockum_ultrastrong_2019, hubener_engineering_2021, genet_inducing_2021, schlawin_cavity_2021}
In these setups, instead of achieving strong modifications of material properties by strong driving, one focuses on realizing strong coupling between light and matter, supported by recent experimental advances. \cite{mavrona_thz_2021, sivarajah_thz-frequency_2019, forn-diaz_ultrastrong_2019, zhang_collective_2016, li_vacuum_2018, keller_few-electron_2017, keller_landau_2020, baranov_ultrastrong_2020, chen_strong_2018}
This might enable the engineering of material properties already with few photons\cite{sentef_quantum_2020, eckhardt_quantum_2021} or even inside a dark cavity, utilizing only the vacuum fluctuations of the light field.\cite{garcia-vidal_manipulating_2021}
Since in this case the material stays in its ground state or energetically close to it, effects from detrimental heating are expected to be reduced and life times to be longer.

In general, photons couple to charged particles or excitations.
Hence there are two main pathways to manipulate electronic properties of a material: either by employing the direct coupling of the light to the electrons; or by utilizing the coupling to other degrees of freedom of the system -- for example lattice vibrations -- that in turn couple to the electrons.
Along the former path several studies have investigated cavity-induced phenomena both from a theoretical and experimental point of view, including superconductivity in which the photons of a cavity provide the pairing glue for the electrons similar to a BCS description,\cite{schlawin_cavity-mediated_2019, gao_photoinduced_2020, chakraborty_long-range_2021, curtis_cavity_2019} suppression of the Drude peak,\cite{rokaj_free_2021, eckhardt_quantum_2021} superradiance\cite{dicke_coherence_1954, kirton_superradiant_2018, ashida_quantum_2020, mazza_superradiant_2019, de_bernardis_cavity_2018, schuler_vacua_2020, guerci_superradiant_2020, skribanowitz_observation_1973, andolina_theory_2020, nataf_rashba_2019} for which it remains an open question whether it can be realized in equilibrium, \cite{rzazewski_phase_1975, bialynicki-birula_no-go_1979, gawedzki_no-go_1981, andolina_cavity_2019, nataf_no-go_2010, stokes_uniqueness_2020} coupling to magnetism\cite{parvini_antiferromagnetic_2020, scalari_ultrastrong_2012, soykal_strong_2010, huebl_high_2013, roman-roche_photon_2021, haigh_triple-resonant_2016, osada_cavity_2016, viola_kusminskiy_coupled_2016} and in particular magnons, \cite{tabuchi_hybridizing_2014, zhang_strongly_2014, goryachev_high-cooperativity_2014, wang_dissipative_2020, zhang_optomagnonic_2016, liu_optomagnonics_2016} excitons\cite{schachenmayer_cavity-enhanced_2015, hagenmuller_cavity-enhanced_2017, deng_exciton-polariton_2010, byrnes_excitonpolariton_2014, schneider_two-dimensional_2018, cortese_excitons_2021, cortese_strong_2019} forming exciton polaritons\cite{novko_cavity_2021, mazza_non-equilibrium_2017} and the modification of topological states of matter. \cite{kibis_band_2011, wang_cavity_2019, appugliese_breakdown_2021, ciuti_cavity-mediated_2021}
Taking a complementary approach, the photons in a cavity also couple to lattice vibrations forming hybrid light-matter excitations -- namely phonon polaritons.
Their potential for steering chemical reactions, \cite{thomas_ground-state_2016, schafer_shining_2021} inducing superconductivity,\cite{thomas_exploring_2019} influencing the ferroelectric phase-transition, \cite{ashida_quantum_2020, latini_ferroelectric_2021}, achieving the redistribution of energy between otherwise non-resonant phonon modes\cite{juraschek_cavity_2021}, or influencing the electron-electron interaction mediated by phonons\cite{sentef_cavity_2018} has recently been investigated.

In this work we explore the possibility of replacing the laser drive for inducing transient superconducting-like states by coupling a material to an optical cavity.
Among the proposals considered to explain the transient states is the suggestion that the laser effectively drives infrared-active (IR-active) phonons in the system that in turn lead to an effective attractive interaction between the electrons through one of two mechanisms put forward.\cite{buzzi_photomolecular_2020, kennes_transient_2017, singla_thz-frequency_2015, kaiser_optical_2014}
These mechanisms involve either a coupling of the phonon coordinate $X$ to the electronic density of the form $X^2 n$, or to the double occupation of the form $X^2 n_\uparrow n_\downarrow$. Such couplings are distinct from the paradigmatic BCS mechanism since the phonons involved are IR-active.
Therefore, a coupling to the electrons proportional to an odd power of the lattice displacement, including a linear coupling as in the BCS mechanism, is in general forbidden by symmetry\cite{kaiser_optical_2014, holstein_studies_1959, hirsch_dynamic_2001} in inversion-symmetric crystals.
In particular in Ref.~\citenum{buzzi_photomolecular_2020} a superconducting response of the system was only observed when driving specific phonon modes in the charge-transfer salt $\kappa$-(BEDT-TTF)$_2$Cu[N(CN)$_2$]Br, abbreviated as $\kappa$-salt from here on out.
Our modelling therefore focuses on this $\kappa$-salt but is kept sufficiently general to be applicable in a broader sense.
Since the considered mechanisms stem from an electron-phonon interaction, we neglect a direct coupling of the electrons to the photons of the cavity.
The remaining coupling between the cavity and the phonons will naturally lead to the formation of phonon polaritons.

After introducing our model we investigate the effect of coupling the phonons in the system to the vacuum fluctuations of a cavity.
For both considered electron-phonon coupling mechanisms this yields an increase of electronic interactions.
This rules out the possibility to enhance $T_c$ in superconductors  by reducing the electron-electron repulsion via vacuum fluctuations through both proposed phonon mechanisms.
Next we consider a weak drive of the cavity, populating the cavity with few, $\mathcal{O}(1)$ photons.
Similar to the case of classical driving of the phonons, this decreases electronic interactions.
We find that in the \ac{sc} but not \ac{usc} regime an increase of the \ac{lmc} does not necessarily lead to a more pronounced effect. Instead a \ac{lmc} that exceeds cavity losses is needed since it determines the time scale on which the photons transfer their energy to the phonons. 
Complementary to this, when increasing the \ac{lmc} to become comparable to the bare cavity frequency hence entering the \ac{usc} regime, we find that driving the emerging phonon polaritons resonantly at their respective eigenfrequency induces an immediate response in the electronic system.
Thus in this \ac{usc} regime a \ac{lmc} that outweighs cavity losses is not strictly required anymore.
Finally, we consider the effects of polariton formation on the electronic spectral function that is in principle observable in an angle-resolved photoemission spectroscopy (ARPES) measurement.
To this end we derive an effective model which we show to capture the dynamics of the system well.
We show a distinctive feature of electrons coupling to polaritons that stem from an IR-active phonon.
The shake-off band\cite{sobota_angle-resolved_2021} that is predicted to appear at a distance from the main spectral peak that equals twice the phonon frequency\cite{sentef_light-enhanced_2017} splits into three bands. We discuss the feasibility of experimentally measuring this feature.

\section{Model}
\label{sec:model}
\begin{figure}[t]
  \centering
    \includegraphics[scale=1.]{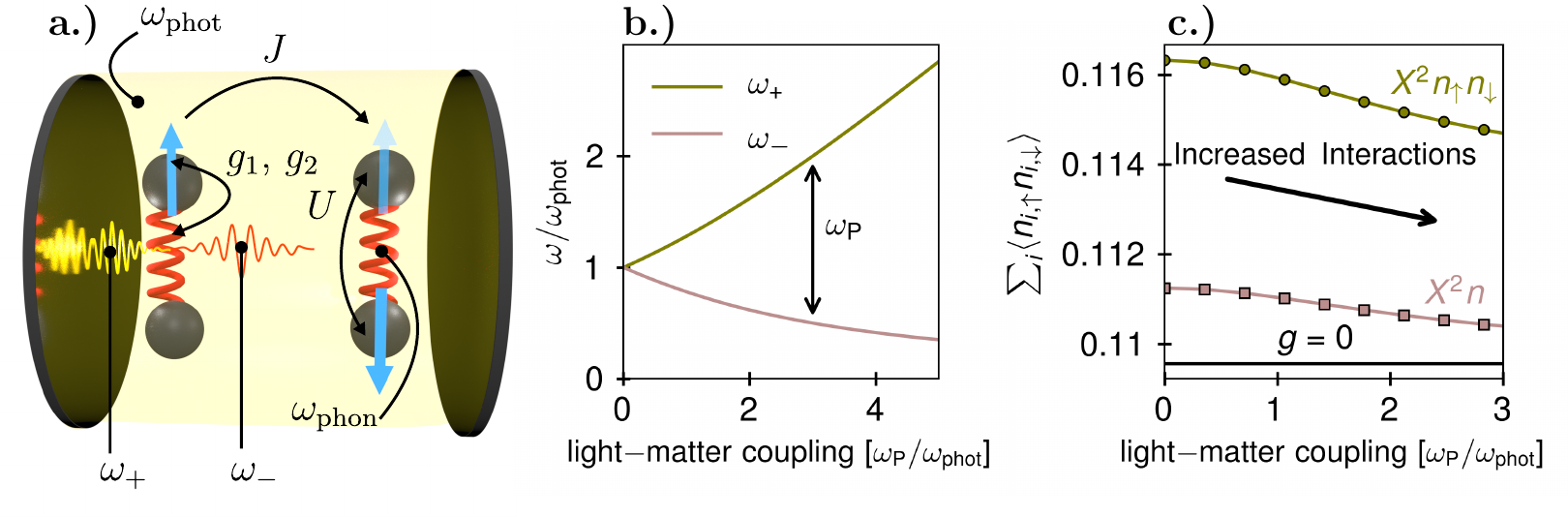}%
  \caption{{\bf Model and effect of dark cavity on electron-electron interactions.} \textbf{a.)} A Hubbard dimer with electrons (blue arrows) coupled to infrared-active phonons (red springs), that are in turn coupled to an optical cavity (black mirror plates). Parameters that appear in the definition of the model in Eq.~(\ref{eq:h-el}), Eq.~(\ref{eq:h-phon}), Eq.~(\ref{eq:h-el-phon}) and Eq.~(\ref{eq:fullH}) are indicated. The frequencies of the phonon-photon hybridized polariton modes $\omega_+$ and $\omega_-$ are indicated.
  \textbf{b.)} Polariton frequencies $\omega_+$ (green curve) and $\omega_-$ (brown curve) for equal bare phonon and photon frequencies.
  \textbf{c.)} Double occupancy Eq.~(\ref{eq:docc}) for different values of the light-matter coupling parametrized by $\omega_{\rm P}$ inside a dark cavity. For both coupling mechanisms $g_1 = 0$, $g_2 = -0.2J$ ($X^2 n_{\uparrow} n_{\downarrow}$, green curve) and $g_1 = 0.5J$, $g_2 = 0$ ($X^2 n$, brown curve) an increase of the light-matter coupling leads to a decrease in double occupancy that is associated with an increase in the effective electron-electron interaction.
  The black line ($g = 0$) marks the value for the uncoupled Hubbard dimer for these parameters.
   We used as a cutoff of the bosonic part of the Hilbert space $N_{\rm B} = 16$.
  Other parameters are the same as in the main text, Sec.~\ref{sec:model}.}
  \label{fig:1}
\end{figure}
We orient our modelling on the $\kappa$-salts discussed in Ref.~\citenum{buzzi_photomolecular_2020} where they were described using a Hubbard model. Other molecular compounds, such as ET-F$_2$TCNQ studied in Refs.~\citenum{kaiser_optical_2014} and \citenum{singla_thz-frequency_2015}, were also found to be described well by a Hubbard model.\cite{okamoto_photoinduced_2007, wall_quantum_2011} 
We therefore also consider a Hubbard model for the matter degrees of freedom, focusing on the two-site version of this model -- the Hubbard dimer.
The Hamiltonian reads
\begin{eqnarray}
\hat{H}_{\rm e^{-}} = - J \sum_{\sigma \in \{\uparrow, \downarrow\}} (\hat{c}_{1, \sigma}^{\dagger} \hat{c}_{2,\sigma} + h.c.) + %
U \sum_{j \in \{1,2\}} \left(\hat{n}^{\rm el}_{j,\uparrow}-\frac{1}{2}\right) \left(\hat{n}^{\rm el}_{j,\downarrow}-\frac{1}{2}\right).
    \label{eq:h-el}
\end{eqnarray}
Here, $\hat{c}_{j, \sigma}$ annihilates -; $\hat{c}_{j, \sigma}^{\dag}$ creates an electron at one of the two sites $j \in \{1, 2\}$ with spin $\sigma \in \{\uparrow, \downarrow\}$.
$J$ denotes the hopping integral, $U$ the onsite repulsion of the electrons and we used $\hat{n}^{\rm el}_{j, \sigma} = \hat{c}_{j, \sigma}^{\dag} \hat{c}_{j, \sigma}$.

We couple each site to an optically active phonon for which the bare Hamiltonian is expressed as
\begin{eqnarray}
    \hat{H}_{\rm{phon}} = \sum_j \omega_{\rm phon} \, \hat{b}_j^{\dagger} \hat{b}_j.
    \label{eq:h-phon} 
\end{eqnarray}
In this expression $\hat{b}_j$ annihilates -; $\hat{b}_j^{\dag}$ creates a phonon with frequency $\omega_{\rm phon}$ at site $j$.
Since the molecules forming the studied solids are centrosymmetric, a coupling between electrons and phonons that is linear in the phonon displacement $\hat{X}_{\text{phon}, j} = \frac{1}{ \sqrt{2 \omega_{\rm phon}}} \left(\hat{b}_j + \hat{b}_j^{\dag} \right)$ is forbidden. \cite{kaiser_optical_2014, holstein_studies_1959, hirsch_dynamic_2001}
The most general term for the electron-phonon interaction where the electrons couple to the quadratic displacement of the phonons reads
\begin{equation}
    \hat{H}_{\rm{phon-e}^{-}} = \sum_j g_1 \left( \hat{b}_j + \hat{b}_j^{\dag} \right)^2 \left( \hat{n}^{\rm el}_{j, \uparrow} + \hat{n}^{\rm el}_{j, \downarrow} \right) +  g_2 \left( \hat{b}_j + \hat{b}_j^{\dagger} \right)^2 \hat{n}^{\rm el}_{j,\uparrow} \hat{n}^{\rm el}_{j,\downarrow}.
    \label{eq:h-el-phon} 
\end{equation}
Here $g_1$ parametrizes the coupling of the phonons to the linear electronic density and $g_2$ that of the phonons to the double occupancy.
In previous works both a coupling that involves a term proportional to the double occupancy\cite{singla_thz-frequency_2015, kaiser_optical_2014} as well as one that only incorporates a coupling to the linear electronic density\cite{kennes_transient_2017} have been considered to understand the optical control of electronic correlations.
In this work we will investigate both mechanisms separately, hence either setting $g_1 \neq 0$ and $g_2 = 0$ or vice versa. 

We model the light degrees of freedom of the optical resonator by a single bosonic mode.
The photon of the cavity is coupled to the optically active phonon whereas its coupling to the electrons is neglected.
Thus we write the total Hamiltonian of the system, including the photon-phonon interaction\cite{mahan_many_2000} and the bare photon energy and collecting the previously defined terms in Eq.~(\ref{eq:h-el}), Eq.~(\ref{eq:h-phon}) and Eq.~(\ref{eq:h-el-phon}) 
\begin{equation}
\begin{aligned}
    \hat{H} =& \hat{H}_{\rm e^{-}} + \hat{H}_{\rm{phon}} + \hat{H}_{\rm{phon-e}^{-}} + \overbrace{\omega_{\rm phot} \, \hat{a}^{\dagger} \hat{a}}^{\hat{H}_{\rm{phot}}}\\
    & +\underbrace{  \sum_j \left[ i\left(\frac{\omega_{\rm{P}} \sqrt{\omega_{\rm{phon}}}}{2 \sqrt{2} \sqrt{\omega_{\rm{phot}}}}\right) \left( \hat{a} + \hat{a}^{\dagger} \right) \left( \hat{b}_j-\hat{b}_j^{\dagger} \right)\right] + \left( \frac{\omega_{\rm P}^2}{4 \, \omega_{\rm{phot}}} \right) \left(\hat{a} + \hat{a}^{\dagger}\right)^2 }_{\hat{H}_{\rm{phon-phot}}}.
    \label{eq:fullH}
\end{aligned}
\end{equation}
Here, $\hat{a}$ annihilates -; $\hat{a}^{\dagger}$ creates a photon in the effective single cavity mode.
$\omega_{\rm phot}$ denotes the bare cavity frequency, $\omega_{\rm P}$ the polariton frequency that parametrizes the phonon-photon or light-matter coupling.
The model is illustrated in Fig.~\ref{fig:1} a.).

The coupling between phonons and photons will lead to the formation of hybrid light-matter states, phonon polaritons.
Their effective frequencies are calculated as\cite{sentef_cavity_2018} (see Appendix \ref{sec:appendix-1-polariton})
\begin{eqnarray} \label{eq:polar_freq}
  \omega_{\pm}^2=\frac{1}{2}\left(\omega_{\rm{phot}}^2+ \omega_{\rm P}^2+ \omega_{\rm phon}^2 \pm \sqrt{\left(\omega_{\rm{phot}}^2+ \omega_{\rm P}^2+ \omega_{\rm phon}^2\right)^2 - 4\omega_{\rm{phot}}^2\omega_{\rm phon}^2} \right).
\end{eqnarray}
For identical phonon and photon frequency, the polariton frequencies are plotted as a function of the coupling $\omega_{\rm P}$ in Fig.~\ref{fig:1} b.).
We call the polariton with the effectively higher frequency $\omega_+$ the upper polariton and that with the effective lower frequency $\omega_-$ the lower polariton.

In what follows the hopping $J$ defines the unit of energy.
For the onsite repulsion $U$ we take an intermediate value of $U = 5J$ that was found in first principles calculations for the $\kappa$-salts.~\cite{buzzi_photomolecular_2020}
The $C-C$ breathing mode of the $\kappa$-salts has an effective frequency of $\omega_{\rm phon}^{\rm eff} \approx 2J$ that is composed the bare phonon-frequency and contributions stemming from the coupling to the electrons.
In Ref.~\citenum{kaiser_optical_2014} it was shown for the molecular compound  ET-F$_2$TCNQ 
that the contribution from the coupling to the electrons can be comparable to or even dominate that from the bare phonon frequency.
We therefore choose parameters such that the two contributions are close to equal in the case of the coupling to the linear electronic density, where we set
\begin{equation}
    g_1  =0.5J \hspace{1mm};\hspace{2mm} g_2 = 0.
    \label{eq:g1_coupling}
\end{equation}
We determine the bare phonon frequency $\omega_{\rm phon}$ such that the effective phonon frequency is equal to the value previously determined for the $\kappa$-salts $\omega_{\rm phon}^{\rm eff} = 2J$.
We find
\begin{equation}
    \omega_{\rm phon} = \left(\sqrt{5} - 1 \right)J \approx 1.24J
\end{equation}
to fulfill this condition. The exact procedure how to obtain the bare phonon frequency is outlined in Appendix \ref{sec:appendix-phonon-freq}.
In the case of coupling exclusively to the double occupancy, the coupling constant $g_2$ is expected to be negative\cite{singla_thz-frequency_2015} for the considered solids.
We anticipate a somewhat smaller absolute value $|g_2| < |g_1|$ compared to the coupling to the linear electronic density (see Eq.~(\ref{eq:g1_coupling})) and thus choose
\begin{equation}
    g_1 = 0 \hspace{1mm};\hspace{2mm} g_2 = -0.2J.
\end{equation}
We note that the detailed values of these couplings do not fundamentally alter our conclusions. Choosing the bare phonon frequency as
\begin{equation}
    \omega_{\rm phon} \approx 2.02J
\end{equation}
creates a resonance of the phonons at frequency $2J$ (also see Appendix \ref{sec:appendix-phonon-freq}).
We couple the phonons resonantly to the cavity and therefore set $\omega_{\rm phot} = \omega_{\rm phon}^{\rm eff} = 2J$.

\section{Electron-Electron Interactions Increase in the Dark Cavity}
\label{sec:gsProperties}
A classical drive of the phonons effectively decreases electron-electron interactions for both electron-phonon coupling mechanisms.\cite{kaiser_optical_2014, singla_thz-frequency_2015, kennes_transient_2017} Here, we investigate the effect that the coupling of vacuum fluctuations of an optical cavity to an optical phonon have on the effective electron-electron repulsion.
As a measure for the repulsion we compute the electronic double occupancy
\begin{equation}
    D = \langle \hat{D} \rangle = \langle \hat{D}_1 + \hat{D}_2 \rangle = \langle \hat{n}_{1, \uparrow}^{\rm el} \hat{n}_{1, \downarrow}^{\rm el} + \hat{n}_{2, \uparrow}^{\rm el} \hat{n}_{2, \downarrow}^{\rm el}\rangle.
    \label{eq:docc}
\end{equation}
An increase in the double occupancy corresponds to a decrease in electronic interactions according to the notion that electrons repel each other less and vice versa.
We set the temperature to $T = 0$ such that the expectation value in Eq.~(\ref{eq:docc}) is evaluated with respect to the \ac{gs}.
We obtain the \ac{gs} via \ac{ed} introducing a cutoff $N_{\rm B}$ in the bosonic part of the Hilbert space.
This is chosen as $N_{\rm B} = 16$, and we have checked that all results are converged with respect to this cutoff.
A more detailed analysis of the convergence in this parameter can be found in Appendix \ref{sec:convergenceGSProps}.

The results for different values of $\omega_{\rm P}$ are shown in Fig.~\ref{fig:1} c.).
Without a cavity ($\omega_{\rm P} = 0$) the coupling to the phonons leads to a slight increase of the double occupancy for both coupling types -- even without a coherent driving.
The coupling of the cavity, however, reverses this effect and leads to a decrease of the double occupancy.
From this observation one can deduce that the presence of the vacuum fluctuations of the cavity increases electronic interactions for both considered electron-phonon coupling mechanisms.

We also consider the effect of finite temperature on the cavity-induced increase in effective electron-electron interactions discussed above.
For this we calculate the thermal expectation value of the double-occupancy in the canonical ensemble according to
\begin{equation}
    \langle \hat{D} \rangle_{\rm therm} = \frac{1}{Z}\sum_n \langle \psi_n | \hat{D} | \psi_n \rangle e^{-\beta E_n},
    \label{eq:thermalAverage}
\end{equation}
where $Z = \sum_n e^{-\beta E_n}$ is the partition function, $E_n$ is the $n^{\rm th}$ eigenenergy of the system, $| \psi_n \rangle$ the corresponding eigenstate, and $\beta = \frac{1}{k_B T}$ the inverse temperature.
To obtain concrete temperature values we take $J = 80\rm{meV}$, which is the value found in ab initio simulations for the $\kappa$-salts performed in Ref.~\citenum{buzzi_photomolecular_2020}. Since these were performed for a triangular-lattice Hubbard model, this can only give a rough order-of-magnitude scale for the temperatures.

\begin{figure}[t]
  \centering
    \includegraphics[scale=1.]{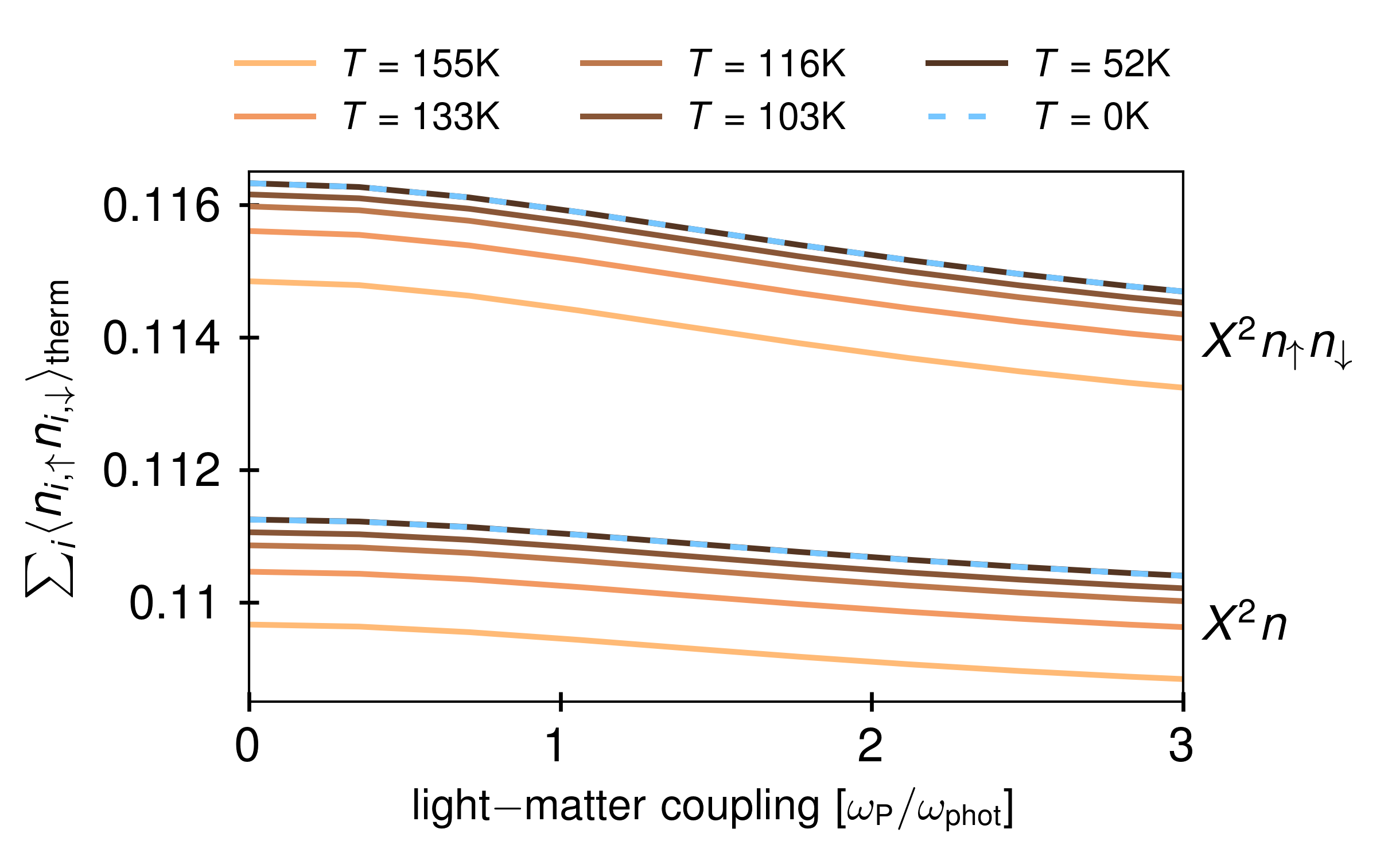}%
  \caption{{\bf Temperature dependence of cavity renormalized double occupancy.}
  Thermal average of the double occupancy according to Eq.~(\ref{eq:thermalAverage}) calculated for different temperatures. We used $J = 80$meV and otherwise the same parameters as in Fig.~\ref{fig:1}c.). The upper curves are calculated with the phonons coupled to the double occupancy of the electrons, while for the lower curves a coupling to the linear electronic density was used, as indicated by the labels.}
  \label{fig:1p5}
\end{figure}

The result are presented in Fig.~\ref{fig:1p5}. Overall, higher temperatures result in a reduction of the double-occupancy but the effect from the coupling to the cavity remains present.

\section{Weak Driving of the Cavity}
\label{sec:timeEvolution}

\begin{figure}[t]
  \centering
    \includegraphics[scale=1.]{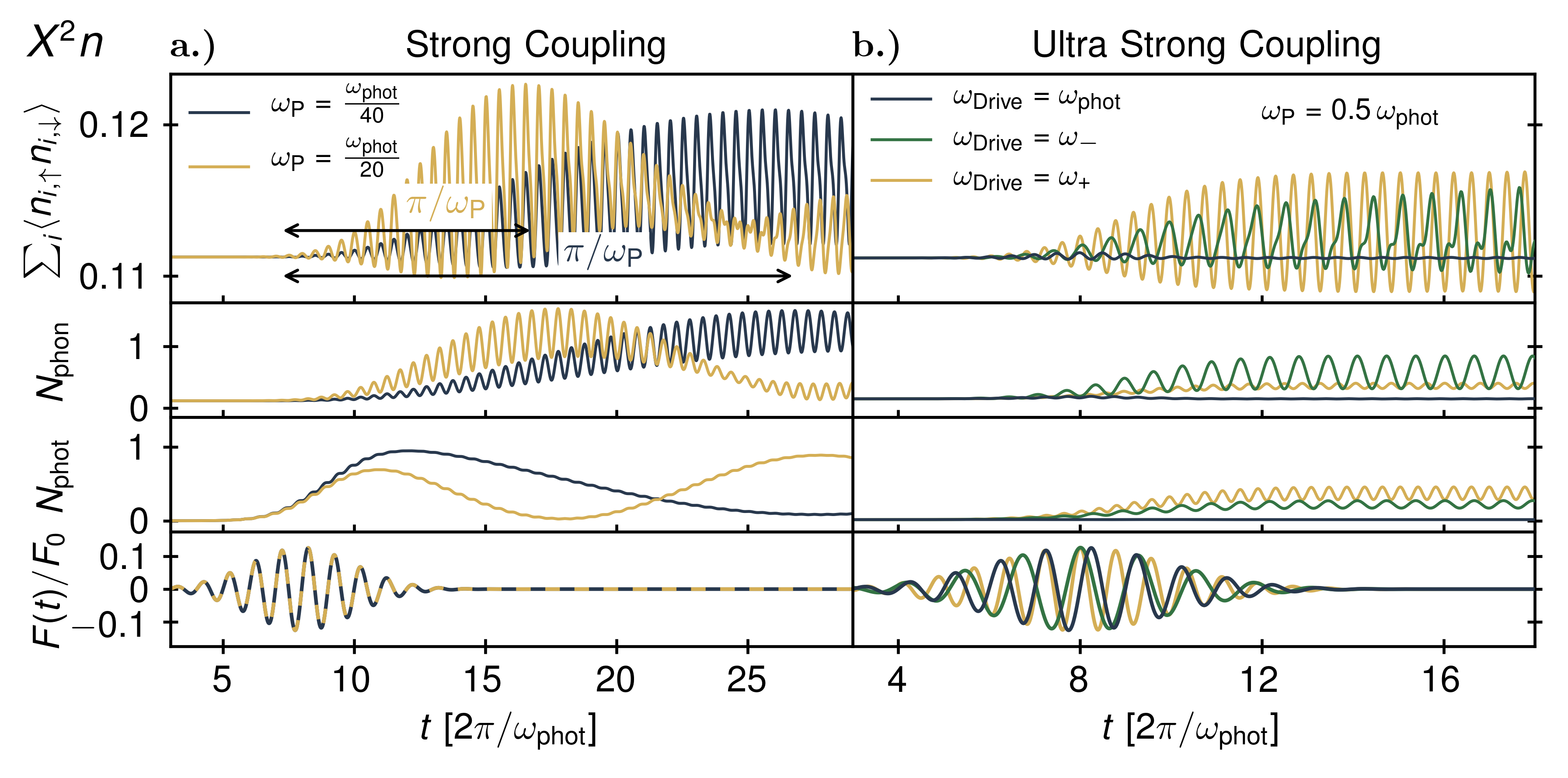}
  \caption{{\bf Driven cavity with coupling of squared phonon displacement to linear electronic density.} Time evolution starting from the \ac{gs} under a coherent drive of the cavity with driving function $F(t)$ (Eq.~(\ref{eq:pump}) shown at the bottom) for a coupling of the phonons to the linear electronic density ($g_1 = 0.5J$, $g_2 = 0$ -- see Eq.~(\ref{eq:h-el-phon})).
  Plotted are the double occupancy of the electrons according to Eq.~(\ref{eq:docc}), the total number of phonons in the system $N_{\rm phon} = \sum_j \hat{b}_j^{\dag} \hat{b}_j$ and the number of photons $N_{\rm phot} = \hat{a}^{\dag} \hat{a}$.
  \textbf{a.)} Strong coupling regime with $\omega_{\rm P} = \frac{\omega_{\rm phot}}{40}$ (blue line) and $\omega_{\rm P} = \frac{\omega_{\rm phot}}{20}$ (yellow line). 
  The photons of the cavity are driven resonantly at $\omega_{\rm Drive} = \omega_{\rm phot}$.
  We observe the beginning of a beating motion between photons and phonons such that it takes the photons a time of $\frac{\pi}{\omega_{\rm P}} = \frac{\tau_{\rm beat}}{4}$ to transfer their energy to the phonons.
  \textbf{b.)} Ultra-strong coupling regime $\omega_{\rm P} = 0.5 \omega_{\rm phot}$.
  The system is driven at three different frequencies: the bare cavity resonance $\omega_{\rm Drive} = \omega_{\rm phot}$ (blue line) and the upper and lower polariton frequency $\omega_{\rm Drive} = \omega_+$ (yellow line) and $\omega_{\rm Drive} = \omega_-$ (green line), respectively.
  When driving the polaritons resonantly both photon number $N_{\rm phot} = \hat{a}^{\dag} \hat{a}$ and the phonon number $N_{\rm phon} = \sum_j \hat{b}_j^{\dag} \hat{b}_j$ increase instantaneously. 
  The cutoff of the bosonic Hilbert space is set to $N_{\rm B} = 10$, and $N_t = 80$ time points per driving period were used.
  }
  \label{fig:2}
\end{figure}
\begin{figure}[t]
  \centering
    \includegraphics[scale=1.]{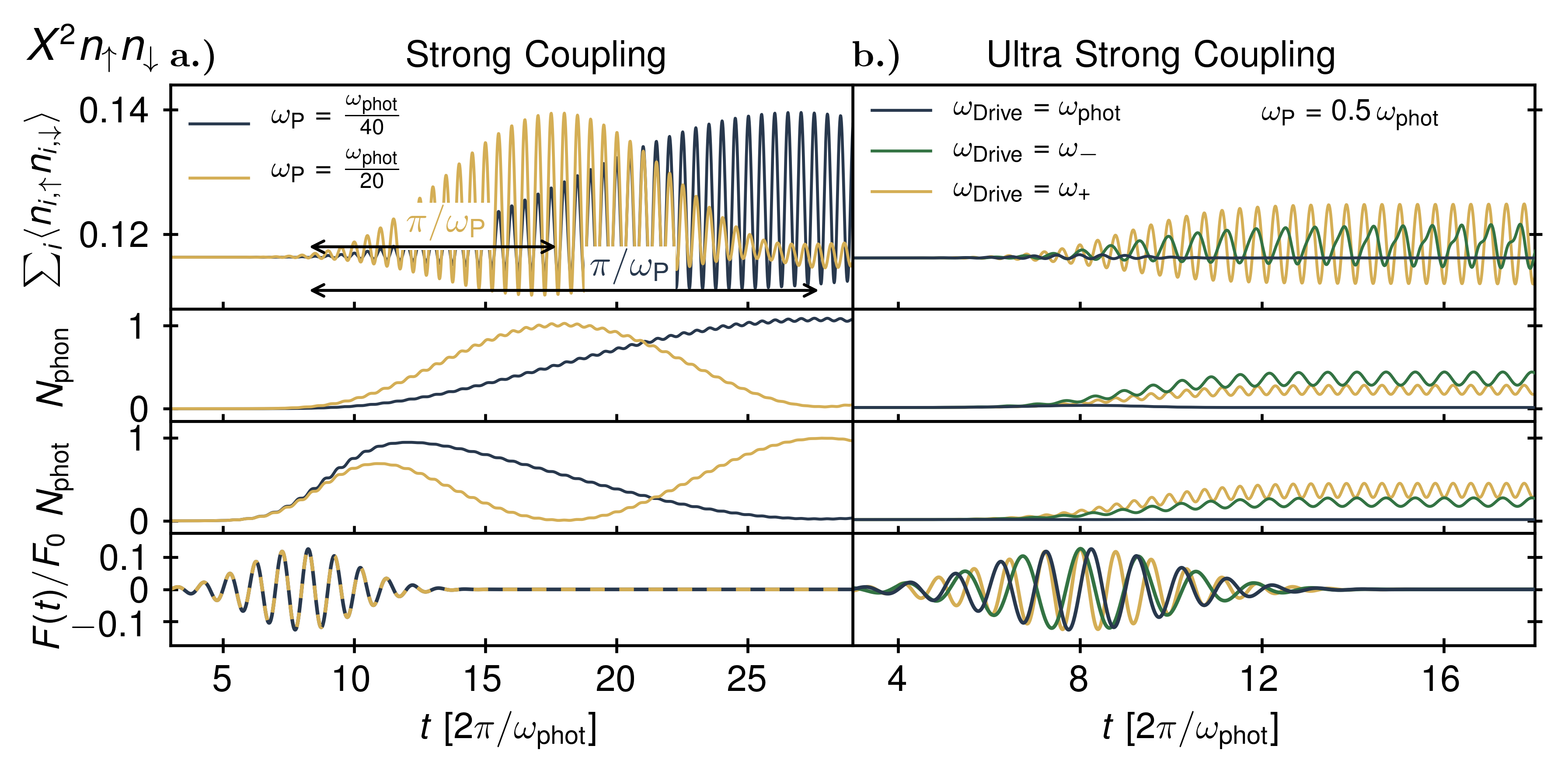}
  \caption{{\bf Driven cavity with coupling of squared phonon displacement to double occupancy.}
  The same setup and identical parameters as in Fig.~\ref{fig:2} but with the phonons coupling to the double occupancy of the electrons ($g_1 = 0$, $g_2 = -0.2J$ -- also see Eq.~(\ref{eq:h-el-phon})) and the bare phonon frequency adjusted such that the effective phonon frequency remains resonant with the cavity as discussed in Sec.~\ref{sec:model}.
  The displayed behaviour is qualitatively similar to the case of the phonons coupling to the local electronic density shown in Fig.~\ref{fig:2}.
  }
  \label{fig:3}
\end{figure}

In this part we apply a weak coherent drive to the cavity and investigate the dynamical change of electronic interactions.
Adding a coherent drive, the time-dependent Hamiltonian reads
\begin{equation}
    \hat{H}(t) = \hat{H} + F(t)\hat{A}_{\rm phot}.
    \label{eq:timeDependentHamiltonian}
\end{equation}
Here, $\hat{H}$ is the Hamiltonian of the undriven system Eq.(\ref{eq:fullH}), $\hat{A}_{\rm phot} = \frac{1}{\sqrt{2 \omega_{\rm phot}}} \left( \hat{a}^{\dag} + \hat{a} \right)$ is the quantized cavity field and $F(t)$ is a pump pulse for which we choose a Gaussian envelope
\begin{equation}
    F(t) = F_0 \frac{1}{\sqrt{2 \pi} \sigma} \exp \left(- \frac{(t - t_0)^2}{2 \sigma^2} \right)  \sin(\omega_{\rm Drive} t).
    \label{eq:pump}
\end{equation}
We use $t_0 = \frac{16 \pi}{\omega_{\rm phot}}$, $\sigma = \frac{4 \pi}{\omega_{\rm phot}}$ and $\frac{F_0}{\sqrt{2 \omega_{\rm phot}}} = \frac{3J}{2}$ as parameters for the driving envelope.
At $t = 0$ the system is prepared in its \ac{gs} and then evolved forward in time via a commutator-free scheme according to Ref.~\citenum{alvermann_high-order_2011}.
Details about the numerical scheme including a convergence study in the finite time-step used as well as the cutoff of the bosonic part of the Hilbert space can be found in Appendix \ref{sec:timeEvolutionAppendix}.

The coupling strength between light and matter inside a cavity is typically classified by comparing it to two distinct quantities: once to the losses of the cavity, where \emph{strong coupling} (SC) refers to a situation in which the coupling exceeds the losses; and once by comparing the coupling to the bare cavity resonance. When the coupling reaches one tenth of the resonance frequency one speaks of \emph{ultra-strong coupling} (USC).\cite{frisk_kockum_ultrastrong_2019}
We do not consider any losses of the cavity, and since we modelled the solid within the cavity with the Hubbard dimer there are no true heating effects either.
We are therefore automatically in the \ac{sc} regime since all time scales are shorter than the (infinite) decay time of the cavity excitation. 
Effects from including a finite cavity life time are discussed later in this section.
Comparing the strength of the \ac{lmc} parametrized in our case by $\omega_{\rm P}$ to the bare cavity resonance we consider two different regimes: two values below \ac{usc} of $\omega_{\rm P} = \frac{\omega_{\rm phot}}{40}$ and $\omega_{\rm P} = \frac{\omega_{\rm phot}}{20}$; and one value within the \ac{usc} regime of $\omega_{\rm P} = \frac{\omega_{\rm phot}}{2}$.

The time evolution of the \ac{gs} of the full coupled system for a coupling of the phonons to the linear electronic density ($g_1 = 0.5J$ and $g_2 = 0$ -- also see Eq.~(\ref{eq:h-el-phon})) is shown in Fig.~\ref{fig:2} and that for the coupling of the phonons to the double occupancy ($g_1 = 0$ and $g_2 = -0.2J$ -- also see Eq.~(\ref{eq:h-el-phon})) of the electrons in Fig.~\ref{fig:3}.
Both coupling mechanisms display qualitatively similar behaviour.
In the case of strong, but not ultra-strong coupling, the pump drives the cavity into an excited state with an increased photon number $N_{\rm phot} = \hat{a}^{\dag} \hat{a}$ within the time duration of the pump.
The strength of the drive is such that only few photons $N_{\rm phot} = \mathcal{O}(1)$ are created.
The energy of the photon excitation is subsequently completely transferred to the phonons on a time scale that is approximately $\frac{\pi}{\omega_{\rm P}}$, as marked in the plot.
When considering even longer times the excitation of the cavity mode and the phonons oscillates back and forth with a period $\tau_{\rm beat} \approx \frac{4 \pi}{\omega_P}$.

In the \ac{usc} case $\omega_{\rm P} = \frac{\omega_{\rm phot}}{2}$, driving the cavity at its bare resonance frequency $\omega_{\rm Drive} = \omega_{\rm phot}$ only yields a weak response.
However, when driving at an increased frequency of $\omega_{\rm Drive} = 2.56J = \omega_+$ that coincides with the upper polariton frequency $\omega_+$ or a decreased frequency of $\omega_{\rm Drive} = 1.56J = \omega_-$ coinciding with the lower polariton frequency $\omega_-$ again a sizeable response is obtained.
In contrast to the \ac{sc} regime, the phonon system reacts immediately in the \ac{usc} regime.
No periodic oscillations between light and matter excitations are observed in this case.
Instead both the phonon number $N_{\rm phon} = \sum_j \hat{b}_j^{\dag} \hat{b}_j$ and the photon number $N_{\rm phot}$ reach a plateau after the drive, with some oscillations on top.

The dynamics of the cavity mode and the phonons can be understood as that of two coupled harmonic oscillators with coupling constant $\omega_{\rm P}$.
To see this we first note that the cavity only couples to the even superposition of the phonon modes on the two sites,
\begin{equation}
\begin{aligned}
  \hat{H}_{\rm phon-phot} &= i \omega_{\rm P} \frac{\sqrt{\omega_{\rm phon}}}{2 \sqrt{2}\sqrt{\omega_{\rm{phot}}}} \left( \hat{a}^{\dag} + \hat{a} \right) \sum_j \left( \hat{b}_j - \hat{b}_j^{\dag} \right) + \frac{\omega_{\rm P}^2}{4 \omega_{\rm phot}} \left( \hat{a}^{\dag} + \hat{a} \right)^2 \\
  &= i \omega_{\rm P} \frac{\sqrt{\omega_{\rm phon}}}{2 \sqrt{2}\sqrt{\omega_{\rm{phot}}}} \left( \hat{a}^{\dag} + \hat{a} \right) \sqrt{2} \left( \hat{b}_0 - \hat{b}_0^{\dag} \right) + \frac{\omega_{\rm P}^2}{4 \omega_{\rm phot}} \left( \hat{a}^{\dag} + \hat{a} \right)^2,
    \label{eq:h-phon-phot-transformed}
    \end{aligned}
\end{equation}
where we have introduced the even combination of bosonic operators
\begin{equation}
    \hat{b}_0^{(\dag)} = \frac{1}{\sqrt{2}} \left( \hat{b}^{(\dag)}_1 + \hat{b}^{(\dag)}_2 \right),
    \label{eq:evenCombination}
\end{equation}
to which a complementary odd combination exists,
\begin{equation}
    \hat{b}_{\pi}^{(\dag)} = \frac{1}{\sqrt{2}} \left( \hat{b}^{(\dag)}_1 - \hat{b}^{(\dag)}_2 \right).
    \label{eq:oddCombination}
\end{equation}
In the strong, but not ultra-strong, coupling regime the two oscillators are weakly coupled when comparing with their bare frequency $\omega_{\rm P} \ll \omega_{\rm phot}$ and $\omega_{\rm P} \ll \omega_{\rm phon}$.
The drive of the cavity displaces one of the oscillators (the photons) such that in the subsequent coupled motion one observes beats -- a phenomenon well known from classical physics.
The period of these beats is classically expected to be $\tau_{\rm beat} = \frac{4 \pi}{\omega_{\rm P}}$, since $\omega_{\rm P}$ equals the splitting of the two eigenmodes of the system, such that one expects the first maximum in the phonon occupation after a quarter period $\frac{\tau_{\rm beat}}{4} = \frac{\pi}{\omega_{\rm P}}$.
This matches well with the observations in Fig.~\ref{fig:2} and Fig.~\ref{fig:3}.

In the \ac{usc} regime light and matter excitations are completely hybridized forming phonon polaritons: One upper polariton with an increased effective frequency of $\omega_+ \approx 2.56J$; and one lower polariton with a decreased effective frequency of $\omega_- \approx 1.56J$ according to Eq.~(\ref{eq:polar_freq}).
This explains why only a small response is observed when driving the system at its bare resonance frequency $\omega_{\rm phot} = 2J$ -- one simply drives the effective oscillators off-resonantly.
When the polaritons are driven at their true resonances instead, with $\omega_{\rm Drive} = \omega_+$ or $\omega_{\rm Drive} = \omega_-$, both phonon and photon degrees of freedom show an immediate response which is a direct consequence of the hybridization of light and matter degrees of freedom.

In a more realistic setup the cavity-matter system experiences losses, either through imperfect mirrors or heating of the material, that might be parametrized by an energy constant $\gamma_{\rm loss}$.
In the case of smaller \ac{lmc} $\omega_{\rm P} = \frac{\omega_{\rm phot}}{20}$ and $\omega_{\rm P} = \frac{\omega_{\rm phot}}{40}$ the response of the system is only triggered with a certain time delay $\frac{\pi}{\omega_{\rm P}}$. In a realistic setup, in order to get a sizeable effect, one thus need a \ac{lmc} of
\begin{equation}
    \omega_{\rm P} > \gamma_{\rm loss}.
    \label{eq:scCondition}
\end{equation}
This is precisely the definition of the \ac{sc} regime.\cite{frisk_kockum_ultrastrong_2019, de_liberato_virtual_2017}
Only increasing the \ac{lmc} compared to the bare cavity frequency does not necessarily yield a larger effect as becomes apparent both from Fig.~\ref{fig:2} and Fig.~\ref{fig:3}.
The comparison of the \ac{lmc} to the cavity losses is therefore the more relevant one in this regime.

In the \ac{usc} regime the response is immediate.
One therefore does not need $\omega_{\rm P} > \gamma_{\rm loss}$ but the challenge lies in reaching a \ac{lmc} that is of comparable size to the bare cavity frequency $\omega_{\rm P} \approx \omega_{\rm phot}$.
This is, in turn, precisely the definition of \ac{usc}\cite{frisk_kockum_ultrastrong_2019, de_liberato_virtual_2017}, which is in particular not a subset of \ac{sc}.

Both electron-phonon coupling mechanisms display qualitatively similar dynamics. In Appendix \ref{sec:classicalDrive} we compare again both mechanisms also for classically driven phonons, finding essentially similar behaviour.

\section{Signatures of electron-polariton coupling in single-particle spectra}
\label{sec:pesWithLMC}
\begin{figure}[t]
  \centering
  \includegraphics[ width=1\textwidth]{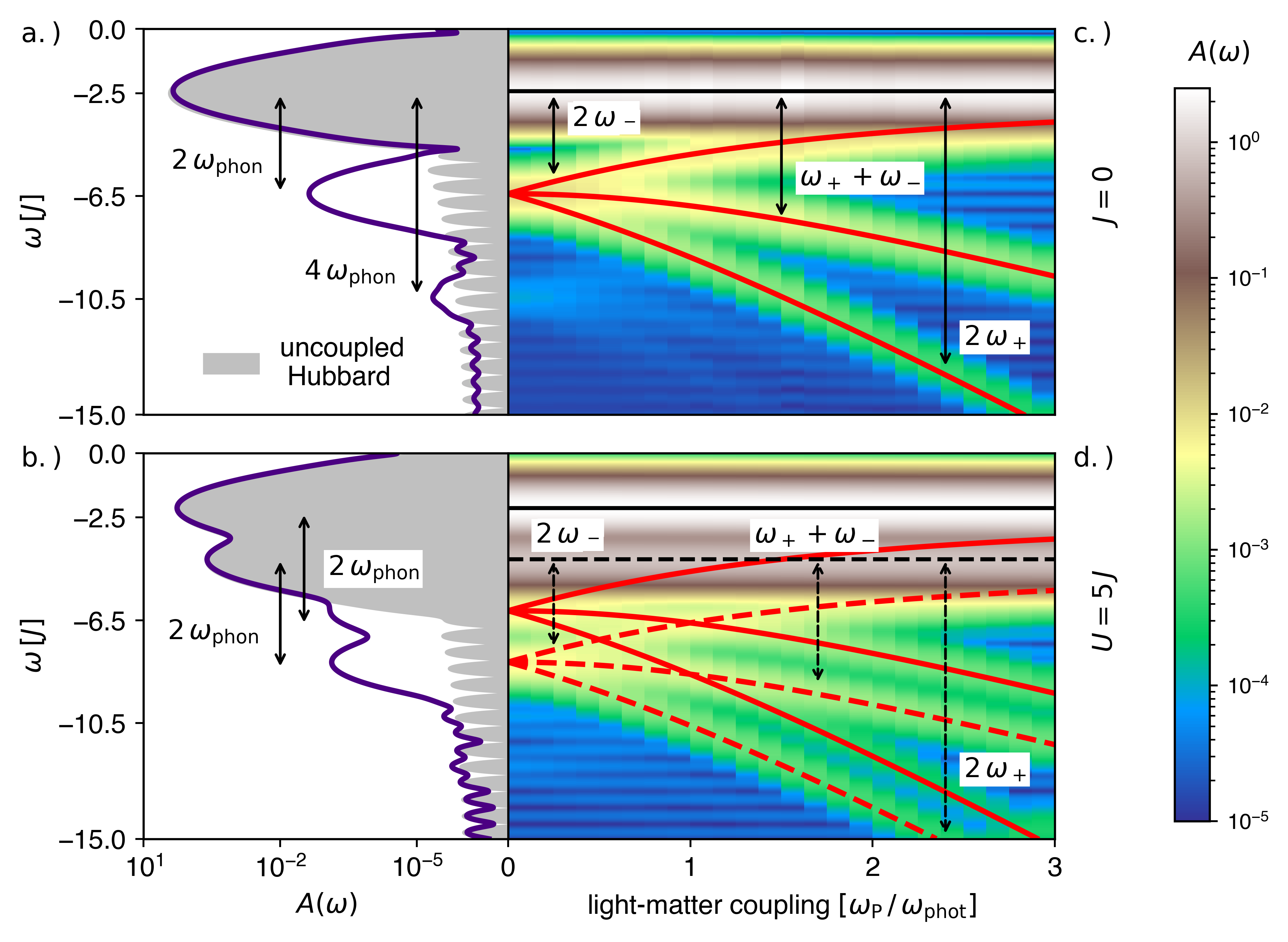}
  \caption{{\bf Effects of polariton formation on spectral functions.} Different spectra according to Eq.~(\ref{eq:ARPES}). Only $\omega < 0$ is shown due to particle-hole symmetry. \textbf{a.)} The $J = 0$ limit without cavity $\omega_{\rm P}=0$. The spectrum of the bare Hubbard dimer is shown in gray. Two side peaks at $2 \, \omega_{\rm phon}$ and $4\, \omega_{\rm phon}$ distance from the main peak are visible.
  \textbf{b.)} Same as a.) for intermediate coupling $U = 5J$ displaying the same phenomenology of $2\omega_{\rm phon}$ side peaks.
  \textbf{c.)} Spectral function in the strong coupling limit $J = 0$ for different values of the light-matter coupling $\omega_{\rm P}$.
  The red lines indicate distances from the lower Hubbard band, $2\omega_+$, $2\omega_-$ and $\omega_+ + \omega_-$.
  At $\omega_{\rm P} = 0$ the spectrum coincides with that shown in a.) whereas upon turning on the light-matter coupling $\omega_{\rm P} > 0$ the shake-off peak splits into three. The intensity scale is indicated on the right.
  \textbf{d.)} Same as c.) just for intermediate coupling $U=5J$ and the case $\omega_{\rm P} = 0$ coinciding with b.). 
  We mark distances from both peaks according to the same combinations of polariton frequencies, $2\omega_+$, $2\omega_-$ and $\omega_+ + \omega_-$, from both bands -- solid lines for distances to the higher lying main band and dashed lines for distances to the lower lying main band.
  Otherwise the same phenomenology as in the strong coupling case, namely the splitting of the shake-off states, is observed.
  $N_{\rm B} = 20$ was used.
  }
  \label{fig:4}
\end{figure}
To further understand the effects of the cavity on electronic properties, we investigate the spectral function for single-particle excitations of the system.
To this end, we focus on the coupling to the quadratic density ($g_1 = 0$, $g_2 \neq 0$) since both coupling mechanisms yield similar \ac{gs} as well as dynamical properties (see Sec.~\ref{sec:gsProperties} and Sec.~\ref{sec:timeEvolution}).
The model including only the coupling to the quadratic density allows for a further simplification when investigating changes induced through the presence of the cavity.
The cavity only couples to the even superposition of the two phonon modes, as already noted in Eq.~(\ref{eq:h-phon-phot-transformed}).
We thus neglect the complementary bosonic mode that is given by the odd combination of phonon excitations.
In Appendix \ref{sec:appendix-1-phonon} we show that this leads to an effective model in which a single boson couples to the total double occupancy of the electrons $\hat{D}$ Eq.~(\ref{eq:docc}) with decreased strength $\tilde{g}_2 = 1 / 2 \, g_2$ and the same coupling to the cavity mode $\tilde{\omega}_{\rm P} = \omega_{\rm P}$.
The dynamics induced through the cavity in this model is qualitatively the same as that in the full model Eq.~(\ref{eq:fullH}).
Additionally, we restore the particle-hole symmetry of the system by writing the electron-phonon coupling as
\begin{equation}
    \hat{H}_{\rm phon-e^{-}} = \tilde{g}_2 \left(\hat{b}_+^{\dag} + \hat{b}_+ \right)^2 \sum_j \left(\hat{n}^{\rm el}_{j \uparrow} - \frac{1}{2} \right)\left(\hat{n}^{\rm el}_{j \downarrow} - \frac{1}{2} \right).
    \label{eq:h-el-phon-new}
\end{equation}
The operator added in the coupling to the electrons in this way only acts like the identity on the electronic part of the Hilbert space and is therefore expected to not change the dynamics.

We calculate the spectral function from the time-evolved states according to the general formalism of time-resolved photoemission spectroscopy,\cite{freericks_theoretical_2009}
\begin{equation}
\begin{aligned}
 A(\omega,t_0)&= \rm{Re} \int dt_1 dt_2 \, e^{i\omega (t_1-t_2)} \, s_{t_1,t_2,\tau}(t_0) \\
 &\left[\bra{\Psi(t_2)}\hat{c}^{\dagger}_{1,\uparrow}\mathcal{T} e^{-i \hat{H} (t_2 - t_1)} \, \hat{c}_{1,\uparrow}\ket{\Psi(t_1)} +\bra{\Psi(t_1)}\hat{c}_{1,\uparrow}\mathcal{T} e^{-i\hat{H} (t_1 - t_2)} \, \hat{c}^{\dagger}_{1,\uparrow}\ket{\Psi(t_2)}\right],
\label{eq:ARPES}
\end{aligned}
\end{equation}
\noindent with $\mathcal{T}$ the time ordering operator. \\
The choice of a particular site or spin orientation does not matter due to symmetry.
As we strictly work at zero temperature the expectation value is calculated with the GS of the system $| \psi_{\rm GS} \rangle$, which is determined via \ac{ed}. Here
$s_{t_1,t_2,\tau}(t_0)$ denotes a Gaussian probe pulse defined as
\begin{equation}
s_{t_1,t_2,\sigma}(t_0)=(2\pi^{3/2}\sigma)^{-1} \, \rm{exp}(-(t_1-t_0)^2/(2\sigma^2)) \, \rm{exp}(-(t_2-t_0)^2/2\sigma^2).
\end{equation}
In the following we take $\sigma = 1.6 / J$ and $t_0 = 5 / J$.
As parameters for the model we set $J=1$ and $U = 5J$, take $\omega_{\rm phot} = 2J$ and an electron-phonon coupling of $\tilde{g}_2 = -0.2J$, unless explicitly denoted otherwise.


The spectral functions for the Hubbard dimer coupled to the phonon mode without cavity,\cite{sentef_light-enhanced_2017} $\omega_{\rm P} = 0$, are shown in Fig.~\ref{fig:4} a.) and b.). We first focus on the case without hopping, $J = 0$, Fig.~\ref{fig:4} a.).
By construction the spectrum is particle-hole symmetric which is why we only show the lower part.
The spectrum of the uncoupled Hubbard dimer is shown in gray exhibiting the well-known lower Hubbard band.
When coupling the electrons to the phonons two shake-off bands at distances $2\omega_{\rm phon}$ and $4 \omega_{\rm phon}$ from the main peak emerge.
No side peaks at uneven multiples of the frequency $\omega_{\rm phon}$ are observed, which is a result of the non-linear coupling proportional to the squared phonon displacement $X_{\rm phonon}^2$.\cite{sentef_light-enhanced_2017}
This coupling essentially squeezes the phonon which leads to the response of the system at twice the bare phonon frequency -- a phenomenon that has previously been predicted and measured.\cite{garrett_vacuum_1997, lakehal_detection_2020}
Additionally the bare value of $U$ is slightly modified - the effect is however quite small and can hardly be seen.
The small wiggles in the spectrum are an artefact of the Gaussian probe pulse.

The spectrum in the intermediate coupling case $U = 5J$, Fig.~\ref{fig:4} b.) similarly exhibits side bands at distances that are compatible with multiples of $2 \omega_{\rm phon}$ from the main bands.
In fact, one would expect the shake-off peaks to appear at a slightly different distance due to the effective frequency of the phonons changing upon coupling to the electrons. This change, however, lies within $1\%$ of the bare frequency (also see Sec.~\ref{sec:model}) and can therefore not be discerned in the plot.

Now allowing for a \ac{lmc} larger that zero $\omega_{\rm P} > 0$ the spectral function for the case of vanishing hopping $J = 0$ is shown in Fig.~\ref{fig:4} c.).
For $\omega_{\rm P} = 0$ the spectrum coincides with that shown in Fig.~\ref{fig:4} a.) exhibiting the previously discussed $2\omega_{\rm phon}$ replica band.
Upon turning on the coupling $\omega_{\rm P} > 0$ we observe a split of this shake-off band into three separate peaks.
We mark in the plot distances from the lower Hubbard band that equal combinations of the polariton frequencies $\omega_+$ and $\omega_-$ namely $2\omega_+$, $2\omega_-$ and $\omega_+ + \omega_-$.
These match the positions of all observed peaks well for all considered coupling strengths.

Essentially the same phenomenology is observed in the intermediate coupling $U = 5J$ case, Fig,~\ref{fig:4} d.).
Here the situations complicated by the natural appearance of two peaks in the lower part of the spectrum of the uncoupled Hubbard dimer.
Still one can observe the $2\omega_{\rm phon}$ replica from both peaks and also track their subsequent split-up into three separate peaks. 
We again mark distances to the two main peaks consistent with the same combinations of polariton frequencies $2\omega_+$, $2\omega_-$ and $\omega_+ + \omega_-$ that match the appearing peaks well.

Now we explain the split-up with the formation of polaritons.
The displacement of the phonon $\hat{X}_{\rm phon}$ can be expressed in terms of a linear combination of the displacement of the upper and lower polariton mode, $\hat{X}_{\rm phon} = c_+ \hat{X}_+ + c_- \hat{X}_-$ where $c_+$ and $c_-$ are two real numbers.
Accordingly the quadratic displacement of the phonons that couples to the electrons Eq.~(\ref{eq:h-el-phon}) transforms under a polariton transformation according to
\begin{equation}
    \hat{X}_{\rm phon}^2 = c_+^2 \hat{X}_+^2 + c_-^2 \hat{X}_-^2 + 2 c_+ c_- \hat{X}_+ \hat{X}_-.
    \label{eq:xSquaredInPolaritons}
\end{equation}
We show the details of the transformation of the coupling between electrons and phonons to a coupling between electrons and polaritons in Appendix \ref{sec:appendix-1-polariton}.
The term in the resulting electron-polariton coupling proportional to $\hat{X}_+^2$ generates a replica peak at distance $2\omega_+$, while the term proportional to $\hat{X}_-^2$ the one at distance $2\omega_-$.
Additionally a mixed term proportional to $\hat{X}_+\hat{X}_-$ appears that generates the peak at $\omega_+ + \omega_-$ distance.
Together this explains the splitting of the electronic shake-off peaks as a unique feature of the electrons coupling to the quadratic displacement of an IR-active phonon.

\section{Discussion and Outlook}

In this work we have investigated the effect of phonon polaritons on electronic interactions.
We have considered two distinct coupling mechanisms between electrons of a strongly correlated material and IR-active phonons, which are in turn coupled to an optical resonator.
Our first finding is that the vacuum fluctuations of the cavity increase the effective electron-electron repulsion.
This might open the path to control electronic interactions in a way that is to date only possible in cold-atom systems.\cite{bloch_many-body_2008}
One possible application would be the triggering of a metal-to-insulator transition by \emph{increased} rather than \emph{decreased} electronic correlations.
To date there are several examples of inducing an insulator-to-metal transition by driving.\cite{rini_control_2007, kiryukhin_x-ray-induced_1997, miyano_photoinduced_1997, fiebig_visualization_1998, mitrano_pressure-dependent_2014, okamoto_photoinduced_2007}
In particular a photo-induced insulator-to-metal transition was observed in in the one-dimensional Mott-insulator ET-F$_2$TCNQ\cite{mitrano_pressure-dependent_2014, okamoto_photoinduced_2007} for which the possibility of controlling electronic interactions through driving of an IR-active phonon with a laser has previously been demonstrated.\cite{singla_thz-frequency_2015, kaiser_optical_2014}
Similarly, effectively reduced correlations by electronic screening through laser-induced electronic excitations have been proposed theoretically \cite{tancogne-dejean_ultrafast_2018, golez_dynamics_2019} and reported experimentally \cite{baykusheva_ultrafast_2021, beaulieu_ultrafast_2021}.

By contrast, we predict that coupling an IR-active phonon to the vacuum fluctuations of an optical cavity will increase electronic correlations, with the possibility of inducing a metal-to-insulator transition.
However, more sophisticated calculations are needed to put our prediction on firmer ground.
The effect of taking the thermodynamic limit should be investigated,\cite{eckhardt_quantum_2021, rokaj_free_2021} and a more detailed description of both the material as well as the cavity is needed -- possibly by building on first principles methods that have recently been extended to cavity QED settings.\cite{tokatly_time-dependent_2013, ruggenthaler_quantum-electrodynamical_2014, pellegrini_optimized_2015} 

The range of realistically achievable changes of effective interactions depends on whether one considers a dark or a driven cavity. In a dark cavity, the relevant quantity is the achievable light-matter coupling strength. Provided that light-matter couplings in the ultrastrong-coupling regime can be attained with quantum materials, modifications of effective interactions in the few-percent range appear realistic. The situation is different in driven cavities. For classically driven systems, changes in effective $U$ of up to 10 percent or even more have been estimated. \cite{singla_thz-frequency_2015,buzzi_photomolecular_2020} Similarly large changes are found in our model simulations of a driven cavity. Therefore, we expect that significant light-induced changes (e.g., potentially cavity-induced superconductivity) might be possible in a driven cavity, presumably at laser intensities below the ones required without a cavity.

One question that has motivated our work is whether a cavity and phonon polaritons can be used to decrease electronic interactions to enable light-induced superconductivity in a similar manner as discussed in Ref.~\citenum{buzzi_photomolecular_2020}.
Despite having practically ruled out this possibility using a dark cavity, a decrease of interactions is being achieved when driving the cavity.
We have investigated the behaviour in two distinct regimes: Once in the strong-coupling case where we have found a delayed response of the matter part with a time delay given by $\frac{\pi}{\omega_{\rm P}}$, where $\omega_{\rm P}$ is the splitting of the two polaritons frequencies; and once in the ultrastrong-coupling regime where we have found a prompt response of the matter system, Sec.~\ref{sec:timeEvolution}.
For further investigation one might promote the model for the matter degrees of freedom to a more sophisticated one.
In a first step possibly, one could investigate a one-dimensional chain that would give access to studying the thermodynamic limit. \cite{eckhardt_quantum_2021, rokaj_free_2021}
In order to research such a model for a sufficiently large system, full diagonalization is not feasible in general anymore due to the exponential growth of the computational cost in the system size.
Instead one might revert to dynamical mean-field theory for correlated electron-boson systems \cite{werner_efficient_2007, werner_phonon-enhanced_2013}, tensor-network based methods\cite{eisert_colloquium_2010, sous_phonon-induced_2021}, or the more recently developed methods based on neural network quantum states.\cite{carleo_solving_2017, hofmann_role_2021}
For the model where the IR-active phonon is coupled to the local electronic density introduced in Ref.~\citenum{kennes_transient_2017} a calculation using a 1D chain to model the electronic system as well as a classical drive of the phonons was performed\cite{sous_phonon-induced_2021} using the infinite time-evolving block decimation (iTEBD)\cite{vidal_classical_2007} method.
The authors found quick decoherence of the phonon motion and phonon-induced disorder in the electronic system. No superconductivity was observed.
It would be interesting to investigate whether similar effects can be found when coupling the phonons to an optical cavity.
In a more sophisticated model it would also be interesting to study the effects of heating of the material or a finite cavity life time.
In our work we have found that large light-matter coupling might not be strictly necessary to achieve sizeable effects, but a light-matter coupling that exceeds losses might be sufficient.
Such a strong-coupling regime has already been reached several decades ago\cite{meschede_one-atom_1985, thompson_observation_1992} and can nowadays be realized in different platforms including  array defect cavities\cite{sivarajah_thz-frequency_2019} and semiconductor heterostructure cavities.\cite{li_vacuum_2018}
Recently also another interesting route to enhance superconducting fluctuations through a parametric drive of IR-active phonons -- possibly with the use of an optical cavity -- has been explored.\cite{grankin_enhancement_2021}

Another possible direction is to make a prediction that helps determine which of the two electron-phonon coupling mechanisms investigated in this work is dominant.
In Ref.~\citenum{kaiser_optical_2014, singla_thz-frequency_2015} the observed drop in reflectivity upon laser driving was explained by a coupling that involved the double occupancy of the electrons.
It was, however, later realized in Ref.~\citenum{kennes_transient_2017} and also becomes apparent from the findings in this work, see Appendix \ref{sec:classicalDrive}, that the observations might also be explained by a coupling to the linear electronic density.
Within our model we have not found qualitative differences between the two coupling mechanisms, neither for ground-state properties nor for their dynamical behavior.
It might be necessary to explore larger system sizes and study the effect of the cavity on longer-range correlation functions and instabilities towards ordered phases in order to identify potential clear distinctions between the two coupling mechanisms.

Finally, we have investigated the electronic spectral function.
For this we have focused on the coupling of the phonons to the double occupancy of the electrons and have derived a simplified model displaying qualitatively and quantitatively similar dynamics upon driving the cavity.
We have identified a distinctive feature of the coupling between electrons and phonon polaritons stemming from IR-active phonons, namely the split-up of the observed shake-off bands into three bands.
Such replica bands due to the coupling between electrons and phonons are well-known in the literature.\cite{sobota_angle-resolved_2021} We note that while we have focussed here on a local, on-site photoemission spectrum without momentum resolution, the corresponding shake-off peaks are expected to appear in a similar fashion in a momentum-resolved ARPES spectrum. This is due to the fact that the long-wavelength photons carry zero momentum transfer compared to the size of the electronic Brillouin zone, thus leading to shake-off peaks separately for each electronic momentum (also see Ref.~\citenum{eckhardt_quantum_2021}).
To observe the split-up of the replica band due to the coupling to an optical resonator proposed here, the linewidth needs to be smaller than the splitting.
For broadening stemming from losses intrinsic to the cavity setup this should be well within reach since the necessary condition is simply the strong-coupling condition.
The question is therefore whether it is possible to achieve a sufficiently strong light-matter coupling to induce spectral weight in the polaritonic shake-off bands that can be detected by an ARPES experiment.

\section*{Acknowledgments}
We thank Nicolas Tancogne-Dejean and Rashmi Singla for fruitful discussions.
We thank Damian Hofmann for help with optimizing the computer code.
Financial  support  by  the  DFG through the  Emmy  Noether  program  (SE  2558/2)  is gratefully  acknowledged.  DMK acknowledges funding by the Deutsche Forschungsgemeinschaft (DFG, German Research Foundation) through RTG 1995 and under Germany's Excellence Strategy - Cluster of Excellence Matter and Light for Quantum Computing (ML4Q) EXC 2004/1 - 390534769.
DMK and MAS acknowledge support from the Max Planck-New York City Center for Non-Equilibrium Quantum Phenomena. 

\section*{Data and Code availability}
Code and data are publicly available at \url{https://github.com/ce335805/PolaritonSqueezing/tree/submissionReference}.
\section*{References}

\bibliography{OrganicsCavity}

\newpage

\appendix

\section{Phonon frequency}
\label{sec:appendix-phonon-freq}

In this part we explain how we obtain the bare phonon-frequency that is a parameter in the Hamiltonian in order to obtain an effective phonon-frequency that is resonant with the cavity photon at $\omega_{\rm phot} = 2J$.
We start by considering the coupling of the phonons to the linear electronic density and write the corresponding electron-phonon Hamiltonian as well as the bare phonon Hamiltonian (compare to Eq.~(\ref{eq:h-phon}) and Eq.~(\ref{eq:h-el-phon}) of the main part)
\begin{equation}
    \hat{H}_{\rm{phon}} + \hat{H}_{\rm{phon-e}^{-}} = \sum_j \omega_{\rm phon} \hat{b}_j^{\dag} \hat{b}_j + \sum_j g_1 \left( \hat{b}_j + \hat{b}_j^{\dag} \right)^2 \left( \hat{n}^{\rm el}_{j, \uparrow} + \hat{n}^{\rm el}_{j, \downarrow} \right). 
\end{equation}
Next we introduce canonical coordinates and momenta
\begin{equation}
    \begin{aligned}
      \hat{X}_j &= \frac{1}{\sqrt{2 \omega_{\rm phon}}} \left(  \hat{b}_j + \hat{b}_j^{\dag} \right)\\
      \hat{P}_j &= i\frac{\sqrt{\omega_{\rm phon}}}{\sqrt{2}}\left(  \hat{b}_j^{\dag} - \hat{b}_j \right)
    \end{aligned}
\end{equation}
and by inserting get
\begin{equation}
\hat{H}_{\rm{phon}} + \hat{H}_{\rm{phon-e}^{-}}  = \sum_j \frac{1}{2} \hat{P}_j^2 + \frac{1}{2} \left( \omega_{\rm phon}^2 + 4 g_1 \omega_{\rm phon} \left( \hat{n}^{\rm el}_{j, \uparrow} + \hat{n}^{\rm el}_{j, \downarrow} \right) \right) \hat{X}_j^2.   
\end{equation}
We now define the effective phonon frequency $\omega_{\rm phon}^{\rm eff}$ as the term multiplying the canonical coordinate of the phonons $\hat{X}_j$ without the one-half. This contains an operator acting on the purely electronic part of the Hilbertspace of which we take the average to obtain a meaningful frequency.
We set this equal to the anticipated value of $\omega_{\rm phon}^{\rm eff} = 2J$ and solve for the bare parameter $\omega_{\rm phon}$ in the Hamiltonian

\begin{equation}
\begin{aligned}
    \left(\omega_{\rm phon}^{\rm eff}\right)^2 &:= \langle \omega_{\rm phon}^2 + 4 g_1 \omega_{\rm phon} \left( \hat{n}^{\rm el}_{j, \uparrow} + \hat{n}^{\rm el}_{j, \downarrow} \right) \rangle_{\rm electronic} = \omega_{\rm phon}^2 + 4 g_1 \omega_{\rm phon} \overset{!}{=} 4J^2\\
    &\overset{g_1 = 0.5J}{\Rightarrow} \omega_{\rm phon} = (\sqrt{5} - 1)J \approx 1.24J.
\end{aligned}
\end{equation}

In the same way we determine the bare phonon frequency when coupling to the double occupancy of the electrons where we write for the part of the Hamiltonian only containing phonon and electron degrees of freedom
\begin{equation}
    \hat{H}_{\rm{phon}} + \hat{H}_{\rm{phon-e}^{-}}  = \sum_j \frac{1}{2} \hat{P}_j^2 + \frac{1}{2} \left( \omega_{\rm phon}^2 + 4 g_2 \omega_{\rm phon} \hat{D}_j \right) \hat{X}_j^2.  
\end{equation}
Hence we can obtain the effective phonon frequency as
\begin{equation}
\begin{aligned}
    \left(\omega_{\rm phon}^{\rm eff}\right)^2 &:= \langle \omega_{\rm phon}^2 + 4 g_2 \omega_{\rm phon} \hat{D}_j \rangle_{\rm electronic} = \omega_{\rm phon}^2 + 4 g_2 \omega_{\rm phon} \langle \hat{D}_j \rangle_{\rm electronic} \overset{!}{=} 4J^2\\
    &\overset{g_2 = -0.2J}{\Rightarrow} \omega_{\rm phon} = 0.4J \langle \hat{D}_j \rangle_{\rm electronic} + 2J\sqrt{0.04 \langle \hat{D}_j \rangle_{\rm electronic}^2 + 1}.
\end{aligned}
\end{equation}
We calculate the expectation value of the double occupancy $\langle \hat{D}_j \rangle$ for the uncoupled Hubbard dimer with which we obtain for the bare phonon frequency
\begin{equation}
    \omega_{\rm phon} \approx 2.02J.
\end{equation}
\section{Convergence of ground state properties in boson cutoff}
\label{sec:convergenceGSProps}

\begin{figure}[t]
  \centering
    \includegraphics[scale=1.]{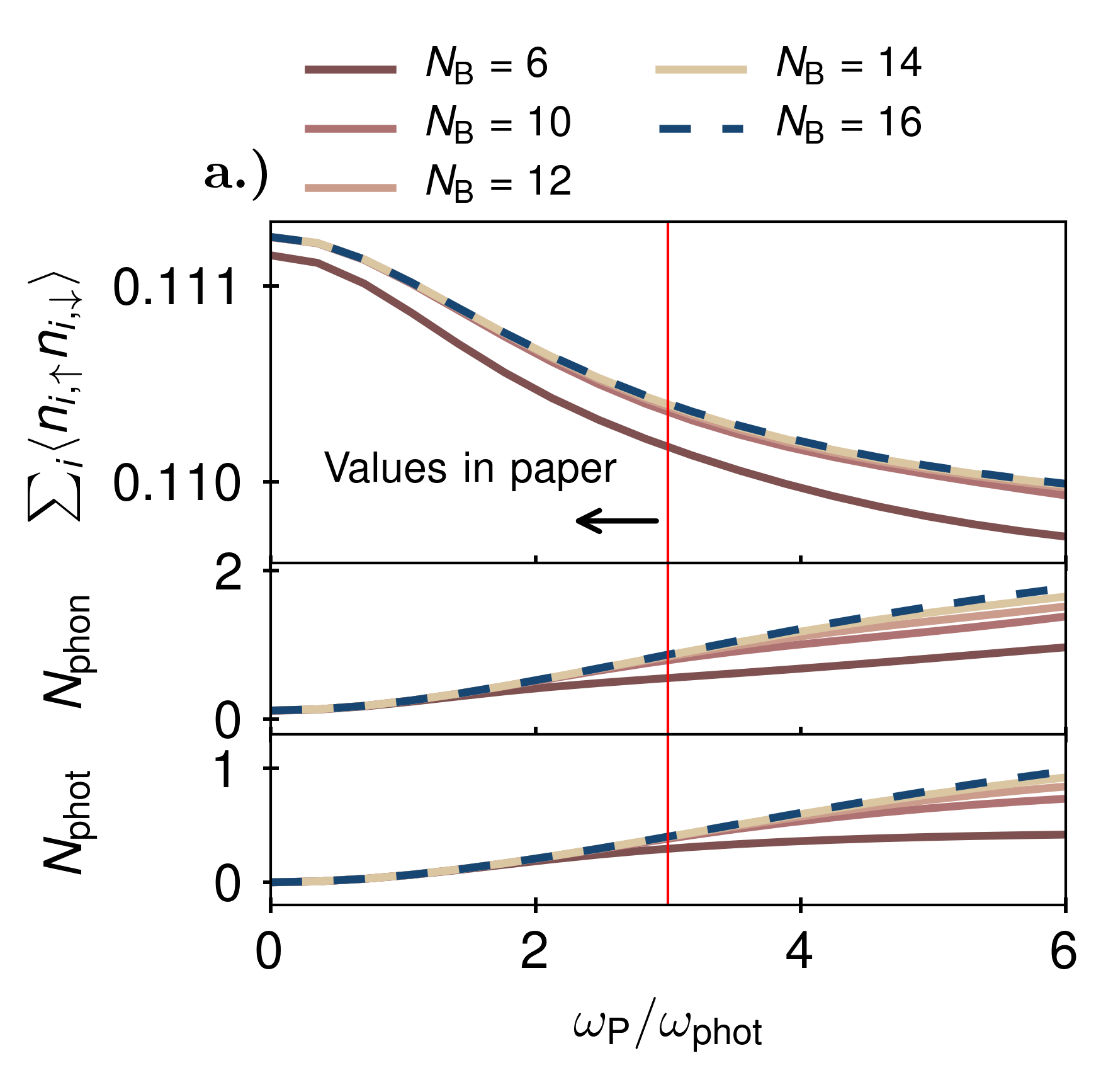}%
    \includegraphics[scale = 1.]{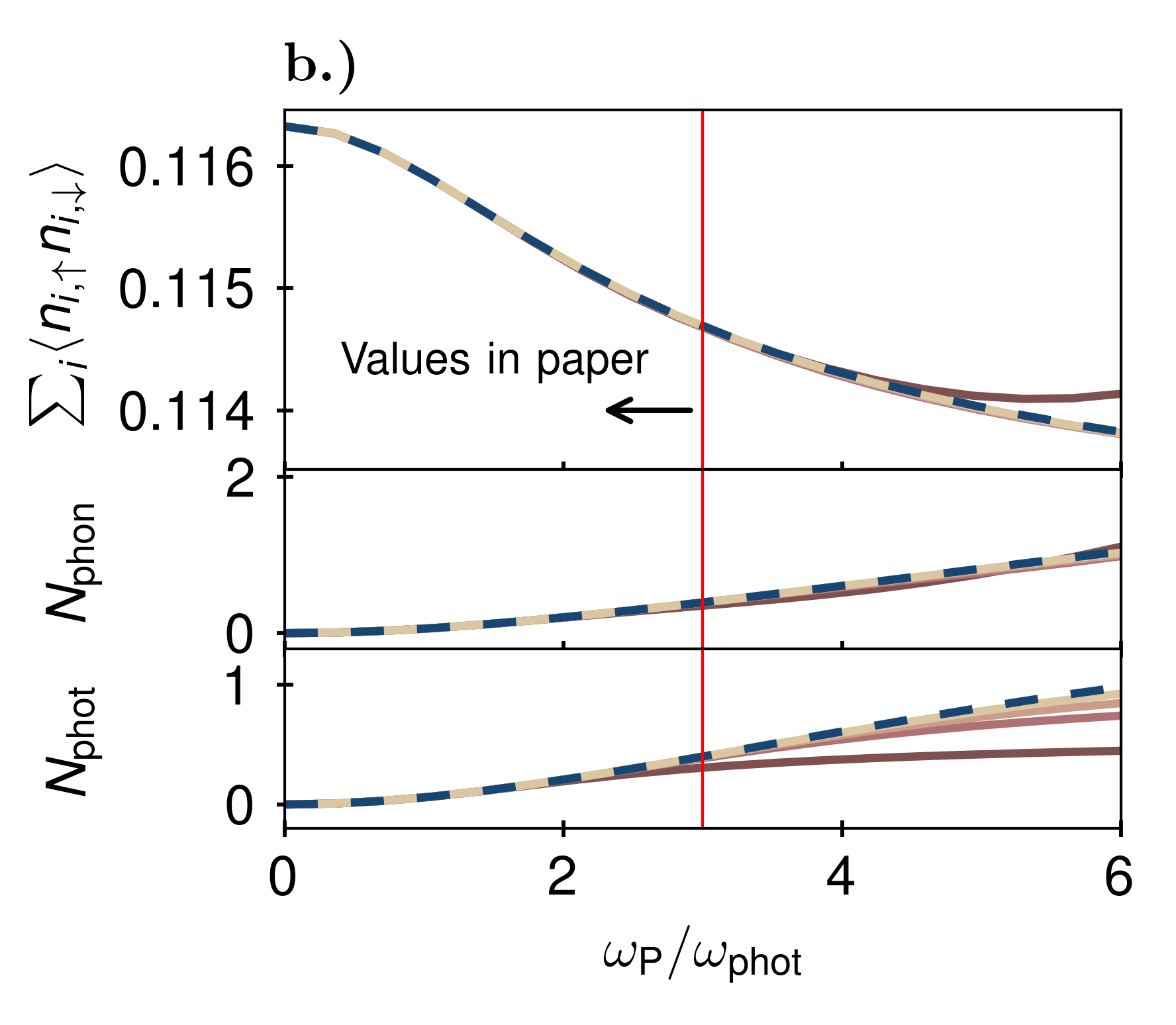}%
  \caption{\textbf{Convergence study of ground-state properties in the cutoff of the bosonic Hilbert space.}
  \textbf{a.)} Double occupancy as a function of the light-matter coupling $\omega_{\rm P}$ calculated with a linear coupling to the electronic density ($g_1 = 0.5$ and $g_2 = 0$ -- see Eq.~(\ref{eq:h-el-phon}) of main part) for different values of the cutoff of the bosonic Hilbert space $N_{\rm B}$. The bosonic cutoff of $N_{\rm B} = 16$ chosen in the paper is sufficient to converge the double occupancy.
  This is further underlined by the boson numbers shown in the lower panel -- once $N_{\rm phon} = \sum_j \hat{b}_j^{\dag}\hat{b}_j$ and once $N_{\rm phot} = \hat{a}^{\dag} \hat{a}$.
  \textbf{b.)} Same convergence study but with a coupling to the quadratic density of the electrons ($g_1 = 0$ and $g_2 = -0.2$ -- see Eq.~(\ref{eq:h-el-phon}) of main part).
  Again, $N_{\rm B} = 16$ is sufficient to converge all data shown in the main part of the paper.
  Other parameters of this convergence study are as in the main text, Sec.~\ref{sec:model}.
  }
  \label{fig:gsPropsConv}
\end{figure}

In this part we check the convergence of the \ac{gs} properties, in particular the double occupancy, in the chosen cutoff of the bosonic Hilbert space $N_{\rm B}$.
We take the same model parameters as in Sec.~\ref{sec:gsProperties} and consider both coupling mechanisms between phonons and electrons discussed in the main part.
The \ac{lmc} is, however, considered up to $\omega_{\rm P} = 6 \, \omega_{\rm phot}$, to show how convergence depends on the coupling strength.
The results are shown in Fig.~\ref{fig:gsPropsConv}.
The double occupancy of the electrons is well converged with a bosonic cutoff of $N_{\rm B} = 16$ as used in the main part for \ac{lmc} of $\omega_{\rm P} \leq 3 \, \omega_{\rm phot}$.
Further beyond that point, around $\omega_{\rm P} \approx 4 \omega_{\rm phot}$ one starts to see deviations between a cutoff of $N_{\rm B} = 16$ and smaller values -- indicating the incomplete convergence at this point.
This result is also further underlined by the number of bosons in the system that remains below $1$ for the values of the \ac{lmc} used in the main part of the paper.

\section{Forward Time Propagation of GS}
\label{sec:timeEvolutionAppendix}

\begin{figure}[t]
  \centering
    \includegraphics[scale=1.]{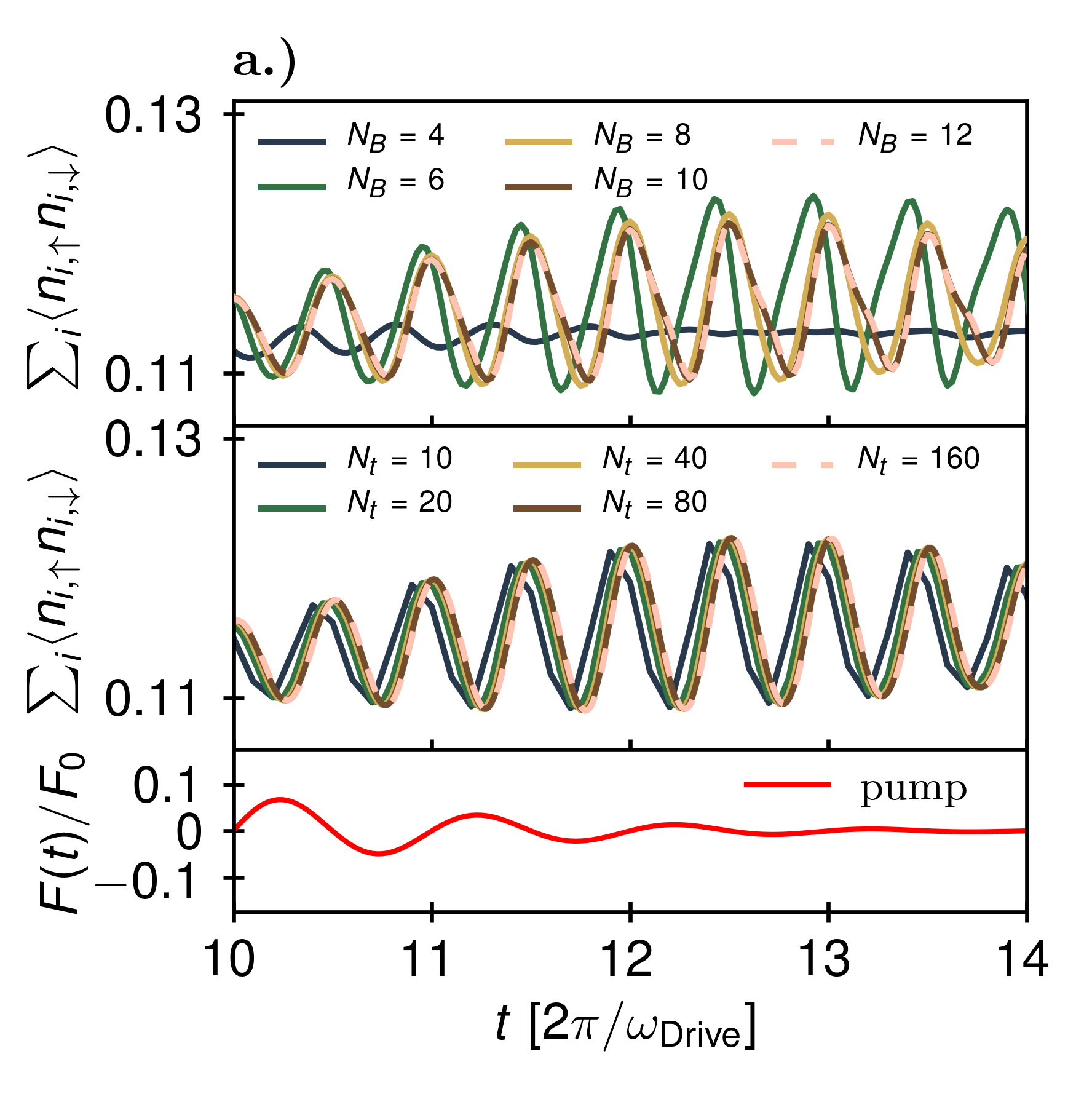}%
    \includegraphics[scale = 1.]{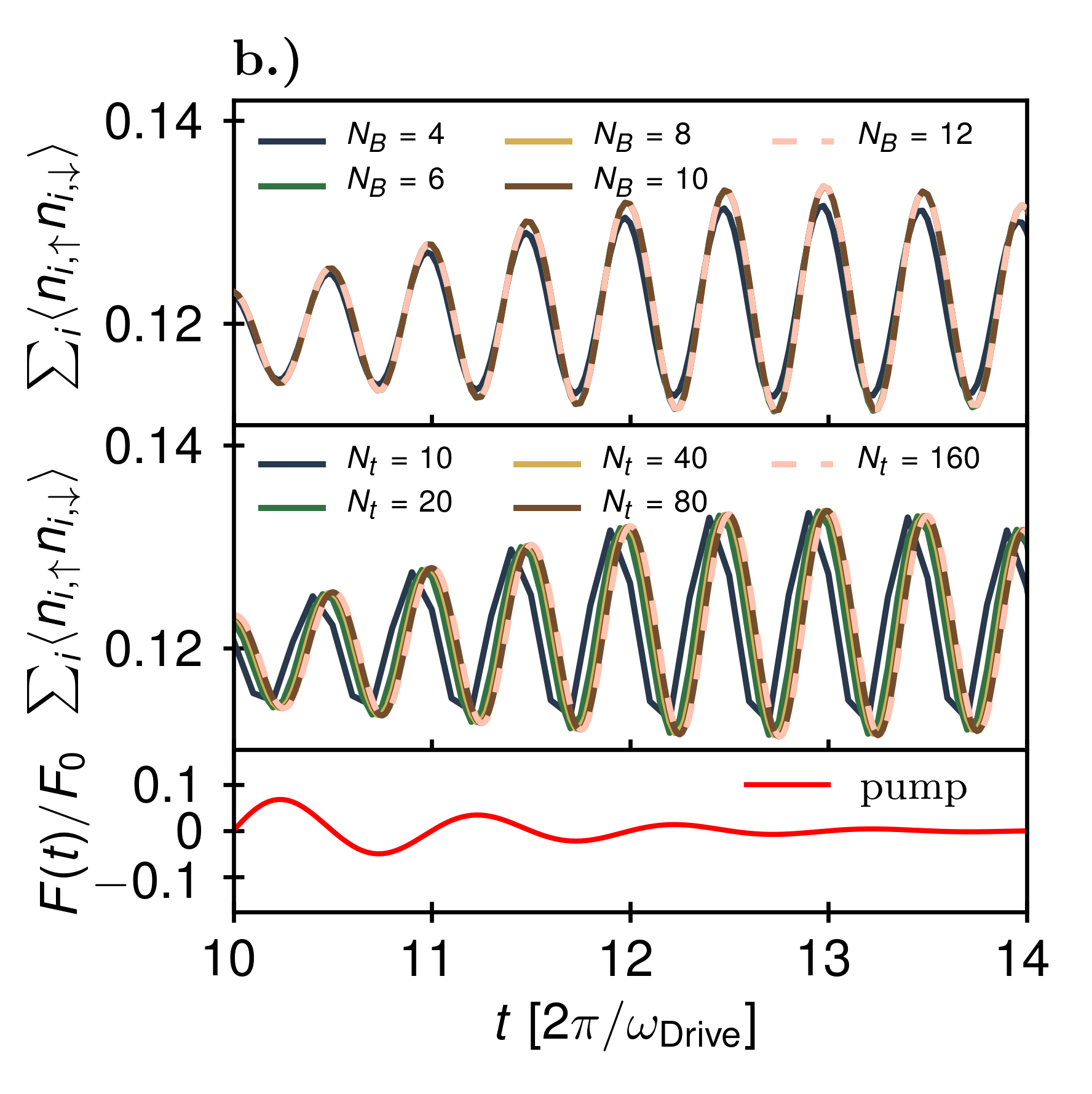}%
    \caption{\textbf{Convergence study of the time-evolved GS under coherent driving of the cavity.}
    \textbf{a.)} Coupling of the squared displacement of the phonons to the linear density of the electrons ($g_1 = 0.5J$ and $g_2 = 0$.)
    Shown is the double occupancy according to Eq.~(\ref{eq:docc}) for different values of the cutoff of the bosonic Hilbert space $N_{\rm B}$ focusing on later times during the pump.
    $N_{\rm B} = 10$ and $N_{\rm B} = 12$ yields almost identical results such that we chose $N_{\rm B} = 10$ for the time-evolution shown in the main part of the paper Sec.~\ref{sec:timeEvolution}.
    $N_t = 40$ and $\omega_{\rm P} = 0.2J$ was used and otherwise the same parameters as in Sec.~\ref{sec:timeEvolution} for this convergence study.
    Additionally the convergence of the double occupancy in the number of time points within one driving period $N_t$ is considered. At $N_t = 80$ the time-evolution is converged with respect to this parameter.
    $N_B = 8$ and $\omega_{\rm P} = 0.2J$ was used and otherwise the same parameters as in Sec.~\ref{sec:timeEvolution}.
    The bottom shows the pump pulse Eq.~(\ref{eq:pump}) as a reference.
    \textbf{b.)} Same convergence analysis for the coupling of the phonons to the double occupancy ($g_1 = 0$, $g_2 = -0.2J$). Otherwise the same parameters were used as in a.).
    The convergence in the bosonic cutoff is better than for the coupling to the linear density. The convergence in the number of time-points per driving period is comparable.
    For the results presented in the main text we use the same value for $N_{\rm B}$ and $N_t$ for both coupling mechanisms.
  }
  \label{fig:timeEvolutionConv}
\end{figure}

In Sec.~\ref{sec:timeEvolution} of the main part we propagate the \ac{gs} of the full coupled system forward in time with the time-dependent Hamiltonian containing an additional coherent drive Eq.~(\ref{eq:timeDependentHamiltonian}) according to
\begin{equation}
     |\psi(t) \rangle = \mathcal{T} e^{-i\int_0^{t} \hat{H}(t') dt'} |\psi_{\rm GS} \rangle
     \label{eq:exactTimeEvolution}
\end{equation}
where $| \psi_{\rm GS} \rangle$ is the \ac{gs} of the system.
We approximate the exact time evolution Eq.~(\ref{eq:exactTimeEvolution}) using finite time-steps $\delta t$ choosing $N_t$ steps within one driving period leading to $\delta t = \frac{2 \pi}{\omega_{\rm Drive} N_t}$.
The time evolution is then computed via the commutator-free scheme introduced in Ref.~\citenum{alvermann_high-order_2011} for a single time-step according to
\begin{equation}
    |\psi(t + \delta t) \rangle = \mathcal{T} e^{-i\int_t^{t + \delta t} \hat{H}(t') dt'} |\psi_{\rm GS} \rangle \approx e^{-i \left( c_1 \hat{H}_1 + c_2 \hat{H}_2 \right)\delta t} e^{-i \left( c_2 \hat{H}_1 + c_1 \hat{H}_2 \right)\delta t}|\psi_{\rm GS} \rangle
    \label{eq:approxTimeEvolution}
\end{equation}
where
\begin{equation}
    c_{1 / 2} = \frac{3 \mp 2 \sqrt{3}}{12} \hspace{1mm} ; \hspace{2mm} \hat{H}_{1 / 2} = \hat{H}\left(t + \left( \frac{1}{2} \mp \frac{\sqrt{3}}{6} \right) \delta t\right).
\end{equation}

We investigate the convergence of this scheme both in the cutoff used for the bosonic part of the Hilbert space $N_{\rm B}$ and the length of the time-step $\delta t$ - or the equivalently the number of time-points within one driving period $N_{t}$.
We consider the double occupancy Eq.~(\ref{eq:docc}) since this is the quantity we are mainly interested in.
Additionally, we found that bosonic quantities like $N_{\rm phot}$ or $N_{\rm phon}$ also considered in the main part usually converge much better with respect to the chosen bosonic cutoff or finite time-step.
We focus on later times at the end of the pump pulse, since an occurring error might build up over time and choose the same parameters as considered in Sec.~\ref{sec:timeEvolution} in the main text.
As for the light-matter coupling we take $\omega_{P} = 0.2J = 0.1 \, \omega_{\rm phot}$ which is on the verge of the \ac{usc} regime.
Overall, we checked that the convergence is similar for different values of the \ac{lmc}.

The results of our analysis are shown in Fig.~\ref{fig:timeEvolutionConv}.
For the convergence in the bosonic cutoff $N_{\rm B}$, results obtained with the coupling to the double occupancy ($g_1 = 0$, $g_2 = -0.2$) are completely converged taking $N_{\rm B} = 6$.
For the coupling to the linear density convergence is significantly slower, however beyond $N_{\rm B} = 8$ changes are relatively small and only quantitative; the qualitative behaviour remains unchanged.
We conclude that a bosonic cutoff of $N_{\rm B} = 10$ is sufficient for our means.

Both coupling mechanisms converge similarly fast in the length of the finite time-step $\delta t$ where we find that $N_t = 80$ steps during a single driving cycle suffice to obtain converged results. 

\section{Comparison Cavity Driving vs. Classical Phonon Driving}
\label{sec:classicalDrive}

\begin{figure}[t]
  \centering
    \includegraphics[scale=1.]{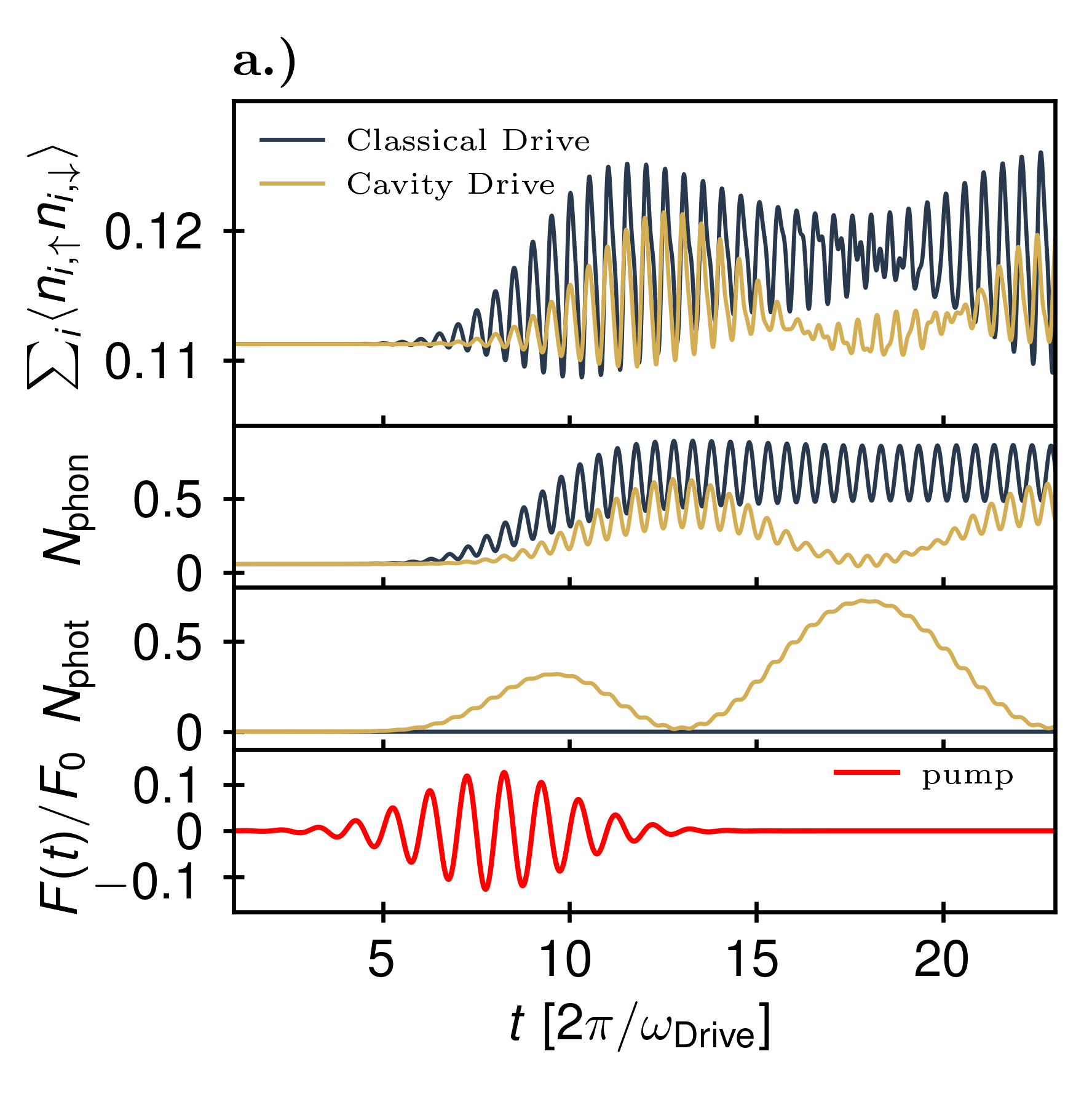}%
    \includegraphics[scale=1.]{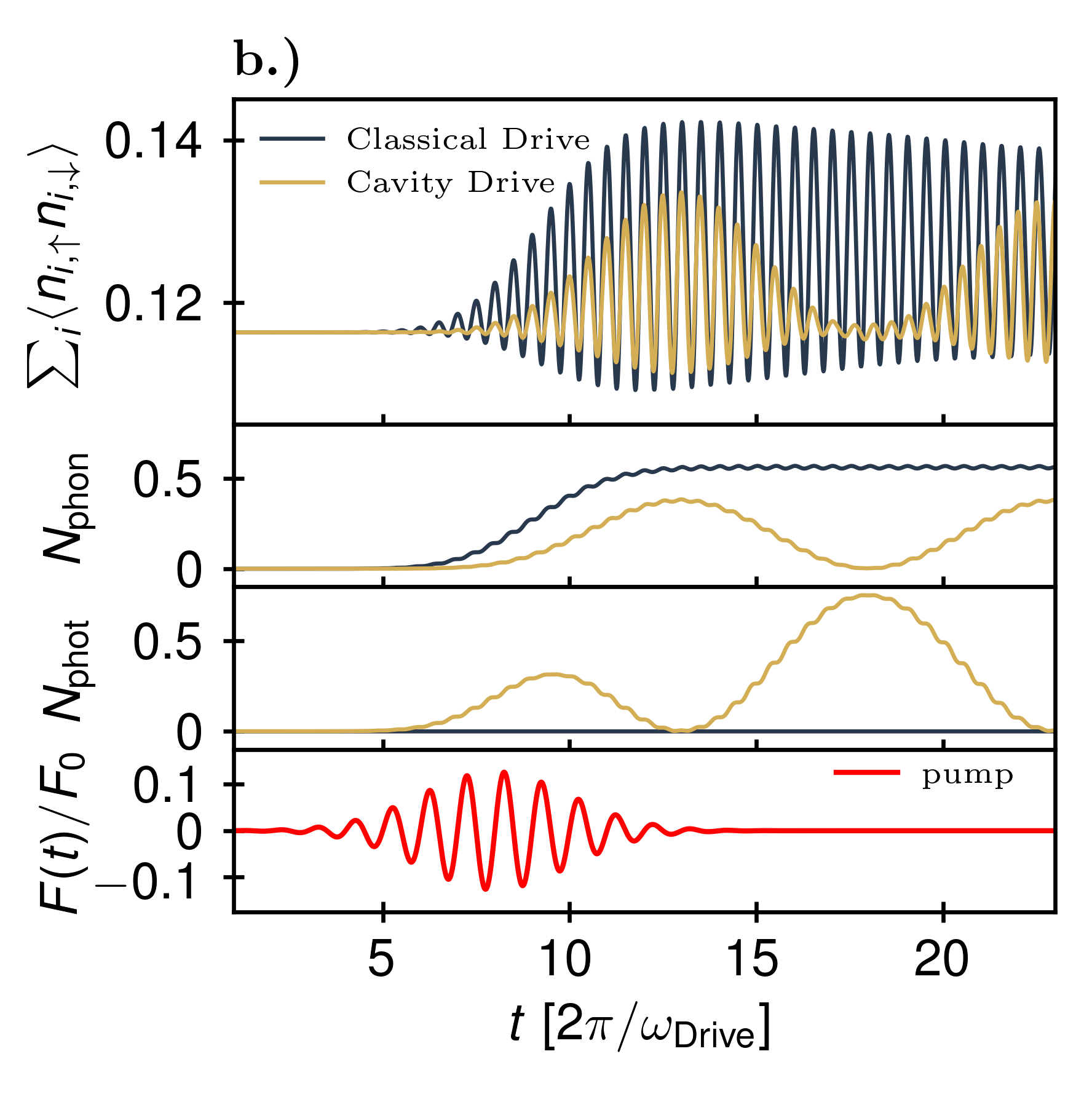}
  \caption{\textbf{Comparison of classical vs. cavity driving.}
  \textbf{a.)} Comparison of the time evolution of the \ac{gs} once driving the cavity and once coherently driving the phonons directly for the linear coupling to the electronic density ($g_1 = 0.5$, $g_2 = 0$ -- see Eq.~(\ref{eq:h-el-phon})).
  Shown are the double occupancy Eq.~(\ref{eq:docc}) at the top, the number of phonons in the system $N_{\rm phon} = \sum_j \hat{b}_j^{\dag}\hat{b}_j$, the number of photons $N_{\rm phot} = \hat{a}^{\dag} \hat{a}$ and the drive-function $F(t)$ according to Eq.~(\ref{eq:pump}).
  Both cases behave similarly with the important difference of oscillations between the excitations of the light field and the phonon mode in the case of driving the cavity.
  We used $\omega_{\rm P} = 0.2J$ and otherwise the same parameters as in Sec.~\ref{sec:timeEvolution}. For the classical drive we set $\omega_{\rm P} = 0$. Otherwise, parameters are the same as for the driven cavity.
  \textbf{b.)} Same comparison between a classical coherent drive of the phonons and the driven cavity for the case of $g_1 = 0$ and $g_2 = -0.2J$ (See Eq.~(\ref{eq:h-el-phon})).
  We monitor the same quantities as in the left plot.
  Again the classical drive of the phonons simply drives them into a coherent state while with a coupled cavity oscillations between excitations of light and matter degrees of freedom are visible.
  Except for the coupling between electrons and phonons we use the same parameters as in the left plot.}
  \label{fig:classicalVSquantum}
\end{figure}

In this part we compare the time evolution of the system for the two different coupling mechanisms between electrons and phonons: once under a classical drive of the phonons and no coupling to the cavity ($\omega_{\rm P} = 0$) and once when coupling to a driven cavity.
This serves the purpose of comparing our results for the driven Cavity to earlier works where a classical drive of the phonons has been considered\cite{kennes_transient_2017, singla_thz-frequency_2015} but also illustrates that both coupling mechanism in fact behave similarly under a classical drive of the phonons.

For the cavity driven system we set the \ac{lmc} to $\omega_{\rm P} = 0.2J$. Otherwise the used parameters are the same as those in Sec.~\ref{sec:timeEvolution} and the time-evolution is calculated in the same way as in that section.
For the classical phonon driving we consider the system uncoupled from the cavity thus setting $\omega_{\rm P} = 0$.
As initial state we take the the \ac{gs} of the system - in this case that without the cavity coupled.
The coherent drive is realized by adding a time-dependent term to the Hamiltonian that reads
\begin{equation}
    \hat{H}_{\rm Drive} = F(t) \frac{1}{\sqrt{2}} \left( \hat{X}_{1, \rm phon} + \hat{X}_{2, \rm phon} \right)
    \label{eq:phononDrive}
\end{equation}
where we have used $\hat{X}_{j, \rm phon} = \frac{1}{\sqrt{2 \omega_{\rm phon}}} \left( \hat{b}_j^{\dag} + \hat{b}_j \right)$.
We drive the system resonantly at the effective phonon frequency $\omega_{\rm D} = 2J$.
$F(t)$ is a Gaussian pulse defined in the main part in Eq.~(\ref{eq:pump}).
The bare strength of the pump $F_0$ cannot be compared directly between the classical and cavity drive since very different matrix elements enter in it: Once the coupling of a drive to the cavity; and once the coupling of the phonons to an external laser.
Since we are interested in a qualitative comparison we simply take the same value for both cases namely $\frac{1}{\sqrt{2 \omega_{\rm phot}}} F_0 = \frac{3J}{2}$.
Otherwise the parameters for the classically driven system are identical to those of the cavity coupled system.
The results for the time-evolution are shown in Fig.~\ref{fig:classicalVSquantum}.

In case of coupling to the double occupancy (shown on the right) the classical drive simply promotes the phonons into a coherent state that oscillates without any damping.
The double occupancy of the electrons also starts oscillating, however, around an average value that is increased from its \ac{gs} value indicating that the drive effectively decreases the electron-electron repulsion. 
The here shown plot can be directly compared to that obtained in Ref.~\cite{buzzi_photomolecular_2020} and Ref.~\cite{singla_thz-frequency_2015} displaying essentially the same phenomenology, albeit without any damping.

In the case of the coupling to the linear density (shown on the left), the coherent state that the phonons are driven into is not as clean as for the quadratic coupling which we attribute to the fact that we here use a larger electron-phonon coupling.
This shows in oscillations in the phonon number, that are larger than in the case of the coupling to the quadratic density.
There also seem to be some overlaying oscillations in the evolution of the double occupancy that are however not reflected in the phonon number.
Whether this is an intrinsic property of this coupling type or some artefact from the model remains unclear at this point.
Qualitatively, the phenomenology between the two couplings is, nevertheless, the same - the driving induces an increased phonon population that in turn leads to an oscillating double-occupancy that is, on average, increased.
A linear coupling of the phonons to the local electronic density might therefore not be ruled out to explain the observations in Ref.~\citenum{singla_thz-frequency_2015} and Ref.~\citenum{kaiser_optical_2014} as was previously noted in Ref.~\citenum{kennes_transient_2017}.

Comparing these results to the cavity driven system, the most prominent difference is an overlaying oscillation between excitations of the phonons and consequently oscillations in the double occupancy; and excitations of the cavity.
This is simply a beating motion of two coupled oscillators after initial displacement of one of the two (the photons in this case).
Otherwise the phenomenology is qualitatively similar.

\section{Approximate One-Phonon model}
\label{sec:appendix-1-phonon}

\begin{figure}[t]
  \centering
    \includegraphics[scale=1.]{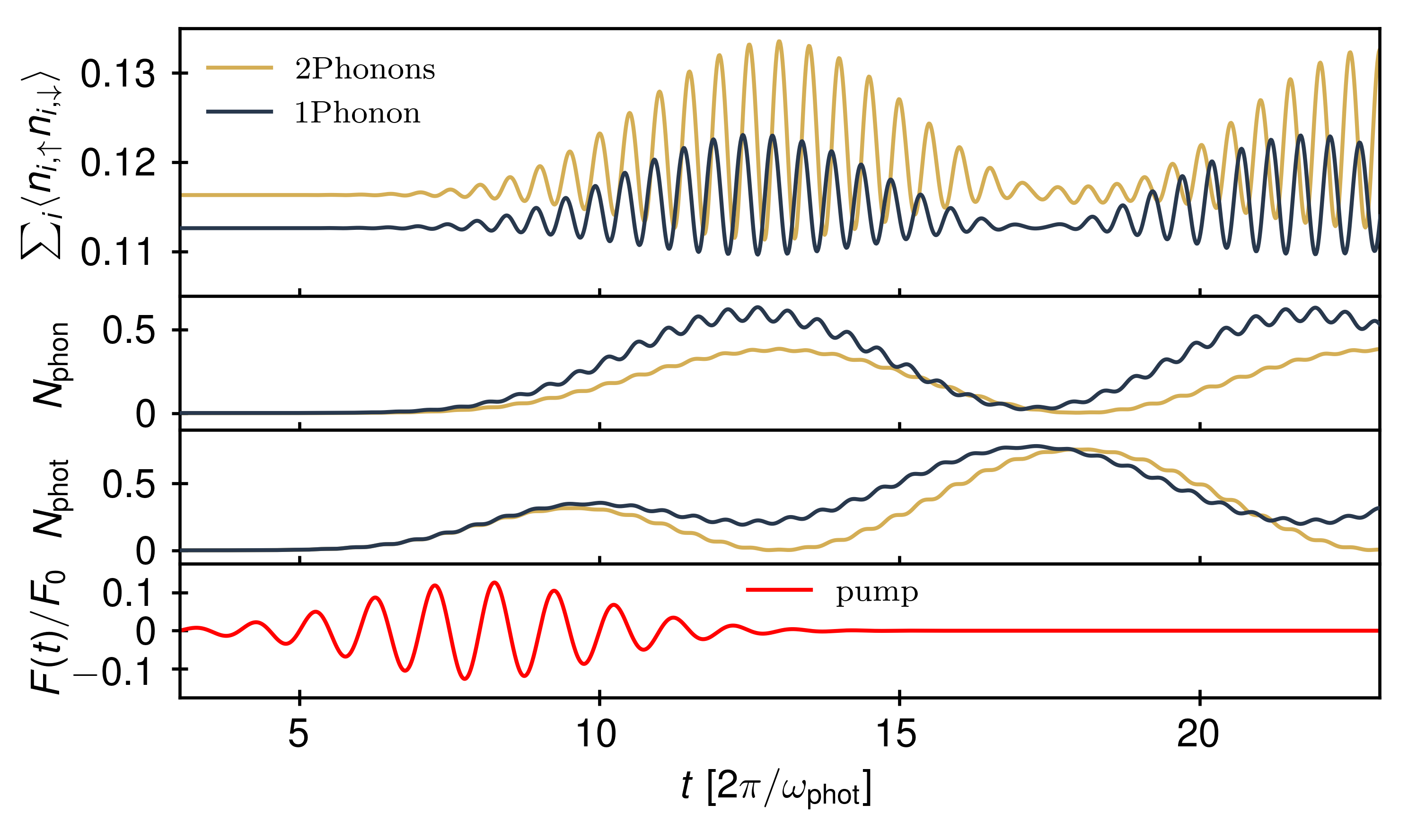}%
  \caption{\textbf{Comparison of One-Phonon to Two-Phonon model.}
  Comparison of the dynamics of the \ac{gs} where phonons couple to the double occupancy ($g_1 = 0$, $g_2 = -0.2J$ - see Eq.~(\ref{eq:h-el-phon})) defined by Eq.~(\ref{eq:fullH}) and the effective one-phonon model defined through Eq.~(\ref{eq:h-el-phon-approx}).
   Shown are the double occupancy Eq.~(\ref{eq:docc}) at the top, the number of phonons in the system $N_{\rm phon} = \sum_j \hat{b}_j^{\dag}\hat{b}_j$, the number of photons $N_{\rm phot} = \hat{a}^{\dag} \hat{a}$ and the drive-function $F(t)$ according to Eq.~(\ref{eq:pump}).
   Under a coherent drive of the cavity both systems undergo similar dynamics.
   The other parameters are as stated in the main text.}
  \label{fig:1PhononVS2Phonon}
\end{figure}

In this part we discuss a simplification of the model in the case of the phonons coupling to the double occupancies of the electrons, i.e., the case of $g_1 = 0$ and $g_2 \neq 0$.
We start by noting that the cavity only couples to the sum of the phonon displacements as already shown in Sec.~\ref{sec:timeEvolution} where we already introduced the even combination of phonons that are annihilated (created) by $\hat{b}_0^{(\dag)}$ Eq.~(\ref{eq:evenCombination}) and odd combinations correspondingly being annihilated (created) by  $\hat{b}_{\pi}^{(\dag)}$ Eq.~(\ref{eq:oddCombination}).
We here reconsider the coupling between electrons and phonons Eq.~(\ref{eq:h-el-phon}) and also write it in terms of the even and odd phonon modes, yielding
\begin{eqnarray}
    \hat{H}_{\rm e-phon} =& \frac{1}{2} g_2 \left( \hat{b}_0^{\dag} + \hat{b}_0\right)^2 \left( \hat{D}_1 + \hat{D}_2 \right) + \frac{1}{2} g_2 \left( \hat{b}_{\pi}^{\dag} + \hat{b}_{\pi}\right)^2 \left( \hat{D}_1 + \hat{D}_2 \right) \\
    &+ g_2 \left( \hat{b}_0^{\dag} + \hat{b}_0\right)\left( \hat{b}_{\pi}^{\dag} + \hat{b}_{\pi}\right) \left( \hat{D}_1 - \hat{D}_2 \right).
\end{eqnarray}
Since Eq.~(\ref{eq:h-phon-phot-transformed}) shows that only the even mode $\hat{b}_0^{(\dag)}$ couples to the cavity and we are mainly interested in dynamics induced through the cavity we neglect the coupling to the odd mode $\hat{b}_{\pi}^{(\dag)}$ and thus approximate the electron-phonon coupling as
\begin{equation}
     \hat{H}_{\rm e-phon} \rightarrow \tilde{\hat{H}}_{\rm e-phon} = \frac{1}{2} g_2 \left( \hat{b}_0^{\dag} + \hat{b}_0\right)^2 \hat{D}.
     \label{eq:h-el-phon-approx}
\end{equation}
This leaves us with a model hosting only a single phonon mode coupling to the cavity with the same strength as before $\tilde{\omega}_{\rm P} = \omega_{\rm P}$ since the additional factor of $\sqrt{2}$ cancels the previously present $\frac{1}{\sqrt{2}}$ term in Eq.~(\ref{eq:fullH}).
This phonon couples to the entire double occupancy of the electrons $\hat{D} = \hat{D}_1 + \hat{D}_2$ with a coupling constant of $\tilde{g}_2 = \frac{1}{2} g_2$.

To show that the approximate model has similar dynamical properties as the original one we again forward propagate the \ac{gs} of each system in time under a coherent driving of the cavity mode.
As parameters we use $g_2 = -0.2J$, $\omega_{\rm P} = 0.2J$ and correspondingly $\tilde{g}_2 = -0.1J$.
We set the frequency of the effective model again such that we expect an effective phonon frequency of $\omega_{\rm phon}^{\rm eff} = 2J$ by choosing
\begin{equation}
    \omega_{\rm phon} \approx 2.01J
    \label{eq:estimateFreqAppendix}
\end{equation}
(also see Appendix \ref{sec:appendix-phonon-freq}).
Otherwise the parameters are as in Sec.~\ref{sec:timeEvolution}.
The result are shown in Fig.~\ref{fig:1PhononVS2Phonon}.

Here, one can see that the dynamics for both models is comparable with two differences.
The effective frequency of the model hosting only a single phonon seems to be slightly higher
than in the model hosting two phonons.
This can be observed in the oscillations of the double occupancy but also in the less complete energy transfer of the cavity to the phonons that indicates that photons and phonons are not quite resonant.
Additionally the beating frequency for the one-phonon model is slightly higher since the frequencies of the effective oscillators lie further apart.
No attempt to correct this slight frequency mismatch was made.
The second difference is the lower double occupancy of the one-phonon model in the \ac{gs} seen at times preceding the pump.
We attribute this to the fact that we dropped some terms in the electron-phonon coupling.

Overall our findings justify the approximation of neglecting the odd mode $\hat{b}_{\pi}^{(\dag)}$ when investigating dynamics induced through the coupling to the cavity with an effective one-mode model.

\section{Polaritonic transformation}
\label{sec:appendix-1-polariton}

We show in this section a basic polaritonic transformation for two coupled oscillators modelling phonon and photon degrees of freedom of a system.
We define canonical position and momentum operators for the photons and phonons
\begin{align}\label{eq:XA}
  \hat{X}_{\rm{phot}} &\equiv \sqrt{\frac{1}{2\omega_{\text{phot}}}}\left(\hat{a}+\hat{a}^{\dagger}\right) \, , \, \, \hat{P}_{\rm{phot}} \equiv -i\sqrt{\frac{\omega_{\text{phot}}}{2}}\left(\hat{a}-\hat{a}^{\dagger}\right)\\
  \label{eq:XB} \hat{X}_{\rm{phon}} &\equiv \sqrt{\frac{1}{2\omega_{\rm{phon}}}}\left(\hat{b}+\hat{b}^{\dagger}\right) \, , \, \, \hat{P}_{\rm{phon}} \equiv -i\sqrt{\frac{\omega_{\rm{phon}}}{2}}\left(\hat{b}-\hat{b}^{\dagger}\right).
\end{align}
The total Hamiltonian of the coupled system can then be rewritten as (also compare with Eq.~(\ref{eq:fullH})):
\begin{align} \label{eq:Hphon_phot_ope}
  {\hat{H}} = \frac{1}{2} {\Big(\hat{P}_{{\rm{phot}}}^2 + \hat{P}_{{\rm{phon}}}^2 + (\omega_{\rm{phot}}^2 + \omega_{\rm{P}}^2)\hat{X}_{{\rm{phot}}}^2 + \omega_{\rm{phon}}^2 \hat{X}_{{\rm{phon}}}^2-2\omega_{\rm{P}}\hat{X}_{{\rm{phot}}}\hat{P}_{{\rm{phon}}}\Big)}
  \end{align}

It is now possible to diagonalize this Hamiltonian defining\cite{sentef_cavity_2018}:
\begin{align} \label{eq:transfo}
  \hat{\tilde{P}}_{{\rm{phon}}} \equiv {\omega_{\rm{phon}}} \, \hat{X}_{{\rm{phon}}} \, , \, \, \hat{\tilde{X}}_{{\rm{phon}}} \equiv -{\omega_{\rm{phon}}}^{-1} \, \hat{P}_{{\rm{phon}}}.
\end{align}
Next we define the angle $\theta$ as
\begin{align}\label{eq:theta}
 &\cos(\theta)=\frac{\Sigma}{\sqrt{1+\Sigma^2}}   \\
  \Sigma &= \frac{\omega_{\text{phot}}^2+\omega_{ \rm P}^2-\omega_{\rm{phon}}^2+\sqrt{(\omega_{\text{phot}}^2+\omega_{\rm {P}}^2+\omega_{\rm{phon}}^2)^2-4\omega_{\text{phot}}^2\omega_{\rm{phon}}^2}}{2\omega_{\rm{phon}}\omega_{ \rm P}}.
\end{align}
To diagonalize the Hamiltonian taking into account the last equations (Eq.~(\ref{eq:transfo}), Eq.~(\ref{eq:theta})), a rotation of $-\theta$ is applied on  $\hat{X}_{{\rm{phot}}}$ and $\hat{\tilde{X}}_{{\rm{phon}}}$ giving respectively $\hat{X}_{+}$ and ${\hat{X}}_-$ - the canonical coordinate operators of the upper and lower polariton respectively. The same transformation is applied to $\hat{P}_{{\rm{phot}}}$ and $\hat{\tilde{P}}_{\rm phon}$ to give $\hat{P}_{+}$ and ${\hat{P}}_{-}$. Performing these transformations, the Hamiltonian can be expressed as:
\begin{align}
    \hat{H} &= \frac{1}{2}{{\begin{pmatrix} \hat{P}_{+} & {\hat{P}}_{-} \end{pmatrix}} {{\begin{pmatrix} 1 & 0 \\ 0 & 1 \end{pmatrix}}} {\begin{pmatrix} \hat{P}_{+} \\ {\hat{P}}_{-} \end{pmatrix}}}
    + \frac{1}{2}{{\begin{pmatrix} \hat{X}_{+} & {\hat{X}}_{-} \end{pmatrix}} {{\begin{pmatrix} \omega_{+}^2 & 0 \\ 0 & \omega_{-}^2 \end{pmatrix}}} {\begin{pmatrix} \hat{X}_{+} \\ {\hat{X}}_{-} \end{pmatrix}}}
\end{align}
The polariton frequencies $\omega_+$ and $\omega_-$ have already been reported in the main text Eq.~(\ref{eq:polar_freq}).
Defining raising and lowering operators for the upper ($\lambda = +$) and lower ($\lambda = -$) in the usual way we can write the diagonalized polariton Hamiltonian as
\begin{align}
    \hat{H}= \sum_{\lambda = \pm}{\omega_{\lambda}\hat{\alpha}^{\dagger}_{\lambda}\hat{\alpha}_{\lambda}}
\end{align}

Using the polariton transformation we can write the coupling term between electrons and phonons Eq.~(\ref{eq:h-el-phon}) with the new $\alpha_{\pm}^{\dagger}$ and $\alpha_{\pm}$ operators:
\begin{equation}
\begin{aligned}
  g_2 \, \left(\hat{b}+\hat{b}^{\dagger}\right)^2 \, &\sum_{i=1,2} \left(\hat{n}^{\rm{el}}_{i,\uparrow}-\frac{1}{2}\right) \left(\hat{n}^{\rm{el}}_{i,\downarrow}-\frac{1}{2}\right) =\\
  &-g_2\left( u_+ \left(\hat{\alpha}_+-\hat{\alpha}_{+}^{\dagger}\right)^2 + u_- \left(\hat{\alpha}_- -\hat{\alpha}_-^{\dagger}\right)^2
  + 2\sqrt{u_+ \, u_-}\left(\hat{\alpha}_+-\hat{\alpha}_+^{\dagger}\right)\left(\hat{\alpha}_- -\hat{\alpha}_-^{\dagger}\right)\right)\\
  &\sum_{i=1,2} \left(\hat{n}^{\rm{el}}_{i,\uparrow}-\frac{1}{2}\right) \left(\hat{n}^{\rm{el}}_{i,\downarrow}-\frac{1}{2}\right).
\end{aligned}
\end{equation}

with  $u_+$ ($u_-$) the phononic contribution of the upper (lower) polariton:
\begin{align}
  u_+ &= \sin^2(\theta) \, \frac{\omega_{+}}{\omega_{\rm{phon}}} \\
  u_- &= \cos^2(\theta)\, \frac{\omega_{-}}{\omega_{\rm{phon}}}
\end{align}
Due to the transformation made in Eq.~(\ref{eq:transfo}) the canonical momenta of the polaritons now couple to the electrons instead of their displacement which is at this point just a matter of definition.
Nevertheless, the bosonic operators appearing here show that both polaritons effectively couple to the electrons explaining the immediate response of the whole system to a drive in the \ac{usc} regime discussed in Sec.~\ref{sec:timeEvolution}.
The presence of three coupling terms with different combinations of bosonic operators also explain the split of the shake-off peak in the electronic spectra into three peaks discussed in Sec.~\ref{sec:pesWithLMC}.
\end{document}